# Measurement of the Josephson Junction Phase Qubits by a Microstrip Resonator

Gennady P. Berman[a], Alan R. Bishop[b], Aleksandr A. Chumak[a,c], Darin Kinion[d], and Vladimir I. Tsifrinovich[e]

[a]Theoretical Division, Los Alamos National Laboratory, Los Alamos, NM 87545, USA
[b]The Associate Directorate for Theory, Simulation & Computation (ADTSC), Los Alamos National Laboratory, Los Alamos, NM 87544, USA
[c]Institute of Physics of the National Academy of Sciences, Pr. Nauki 46, Kiev-28, MSP 03028, Ukraine
[d]Lawrence Livermore Laboratory, Livermore, CA 94551, USA
[e]Polytechnic Institute of NYU, 6 MetroTech Center, Brooklyn, NY 11201, USA

**Abstract**

The process of measurement of a phase qubit by a resonant microwave cavity is considered for various interactions between the qubit and the cavity. A novel quasiclassical approach is described based on adiabatic reversals of the qubit state by an effective field. A similar approach was implemented earlier for the detection of electron and nuclear spins using magnetic resonance force microscopy (MRFM), but this approach has not previously been used for the measurement of a quantum state. Quasiclassical and quantum regimes are described. We consider both linear and nonlinear resonators. The effects of the environment on the process of measurement are also analyzed.

## I. Introduction

Superconducting Josephson junctions are now considered as the most realistic candidates for solid state qubit implementation. (See, for example, Refs. [1-4].) The research in this field is mainly concentrated on (i) manipulation with one and two qubits, (ii) measurement of qubit states, and (iii) improving the performance of qubits by reducing the effects of decoherence and relaxation. In this paper, we consider the measurement of phase qubits using a resonant microstrip resonator. Generally, the measurement of the qubit state can be performed by (i) measurement of the probabilities of the basis states of a qubit and (ii) implementing a state tomography when the phase information can also be extracted [1-4]. Usually, the procedure of measurement depends on both the type of qubits to be measured and the measurement device. In the scheme considered in this paper we use a so-called flux biased phase qubit (see below), and as a measurement device we use a microstrip resonator (which is termed a resonator, a cavity, or an oscillator). Usually one considers the shift of the frequency of the resonator which results from the interaction between a qubit and a resonator, and which depends on the state of the qubit [5]. Usually, this shift is small enough, and can be calculated by using a regular perturbation approach based on the renormalization of the unperturbed eigenvalues and eigenfunctions [5].

In this paper, we propose a different approach to the problem of measurement of a qubit state, by using the method of adiabatic reversals of a qubit. We assume that the cavity field is in a quasiclassical state, which is the typical situation with a measurement device. In this approach an adiabatic invariant is created in the total "qubit-cavity" system. The operational conditions are chosen in such a way that the qubit follows the direction of the effective field. The shift of the phase or frequency of the cavity field, which is detected, depends on the direction of the qubit state relative to the direction of the oscillating effective field. We show that this approach gives different results compared with the standard perturbation methods. Namely, the shift of the frequency is first order in the perturbation parameter instead of second order as was found earlier in the regular perturbation approach. In the first part of the paper, we give a detailed description of the phase qubits and their interactions with the resonator, including a nonlinear resonator. In the second part, we present the details of the proposed approach. We consider quasiclassical and quantum approaches as well as the influence of the environment on the process of measurement.

### 1.1. Josephson equations

In 1962 Brian D. Josephson published his theory about properties of weakly coupled superconductors, which later were called Josephson junctions. At high temperatures a Josephson junction behaves like an ohmic resistance. Once superconducting, the insulating barrier is penetrated by the wave functions of both superconductors. The overlapping wave functions allow Cooper pairs to flow without resistance. Josephson predicted that this supercurrent depends on the phase difference, $\delta$, between wave functions in both superconductors. It is given by,

$$I = I_c \sin\delta , \tag{1.1}$$

where $I_c$ is the maximum dissipationless current, which can flow through the junction.

Now suppose that we connect the two superconducting regions to the two terminals of a battery so that there is a potential difference, $V$, across the junction. In this case, the phase difference (in what follows simply phase) varies in time, and its evolution is governed by the equation,

$$\frac{d\delta}{dt} = \frac{2\pi}{\varphi_0} V, \tag{1.2}$$

where $\varphi_0$ is the flux quantum equal to $h/2e$, $h$ and $e$ ($e > 0$ here) are Planck's constant and the electron charge, respectively. The derivation of Eqs. (1.1) and (1.2), which are known as the Josephson equations, can be found, for example, in the book [6].
It follows from Eqs (1.1) and (1.2) that the inductance of the Josephson junction is given by,

$$L_e = L_J / \cos\delta , \tag{1.3}$$

where, $L_J = \varphi_0 / 2\pi I_c$. As we see, the inductance, $L_e$, is a nonlinear element since it depends on $\delta$ and, correspondingly, on the current, $I$.

### 1.2. Current-biased phase qubit

Usually, Josephson junctions can be modeled as a parallel combination of an ohmic resistor, $R$, which is due to the current of normal (not Cooper pairs) electrons, the Josephson element, described by Eq. (1.2), and a capacitor, $C$, accounting for the capacitance of the electrodes. Then the total current is a sum of the three constituents (see Fig. 1),

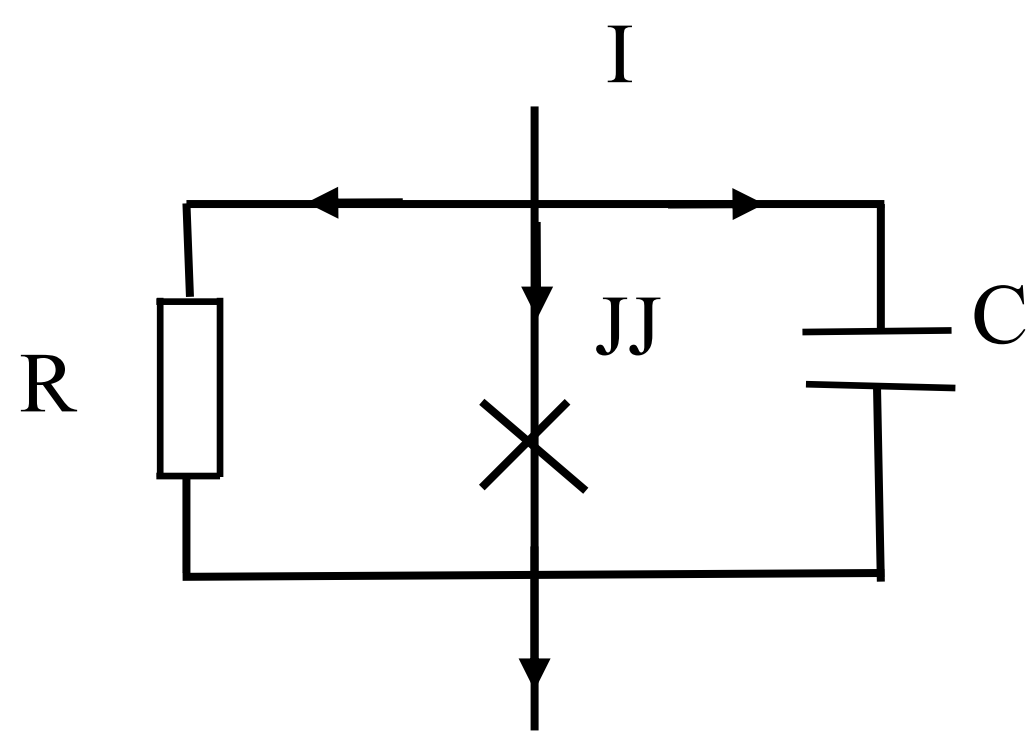

Fig. 1. Current-biased Josephson junction. A resistor, $R$, is due to the presence of normal electrons; a capacity, $C$, accounts for the capacitance of the electrodes; $JJ$ regards to the supercurrent element.

$$I = I_c \sin\delta + \frac{V}{R} + C\frac{dV}{dt} = I_c \sin\delta + \frac{\varphi_0}{2\pi R}\dot{\delta} + \frac{C\varphi_0}{2\pi}\ddot{\delta}\,. \tag{1.4}$$

Multiplying Eqs. (1.4) by $\frac{\varphi_0}{2\pi}$, we obtain the evolution equation for $\delta$ in the form,

$$m\ddot{\delta} + \frac{m}{RC}\dot{\delta} + \frac{\partial U(\delta)}{\partial \delta} = 0\,, \tag{1.5}$$

where the "mass" $m$ is given by $m = C\left(\frac{\varphi_0}{2\pi}\right)^2$, and the "potential energy", $U(\delta)$, is

$$U(\delta) = -E_J\left(\frac{I}{I_c}\delta + \cos\delta\right), \tag{1.6}$$

and the parameter, $E_J$, is equal to $I_c\varphi_0/2\pi$. Note, that the potential energy of the supercurrent element can be obtained as the energy required to form the state with a given value of the supercurrent. It is given by,

$$\int^{t} I_J(t')V(t')dt' = \frac{\varphi_0}{2\pi}\int^{t} I_J(t')\frac{d\delta(t')}{dt'}dt' = E_J\int^{\delta}\sin\delta' d\delta' = -E_J\cos\delta,$$

where $I_J(t') = I_c\sin\delta(t')$, and Eq. (1.2) was used.

Let us neglect in Eq. (1.5) the term with a resistance, $R$. Then in the absence of dissipation the circuit Lagrangian corresponding to Eq. (1.5) can be written in the form,

$$L = \frac{m}{2}\dot{\delta}^2 - U(\delta). \tag{1.7}$$

It can be easily verified that in this case the Lagrange equation,

$$\frac{d}{dt}\frac{\partial L}{\partial\dot{\delta}} - \frac{\partial L}{\partial\delta} = 0, \tag{1.8}$$

coincides with Eq. (1.5). The canonical momentum, $q$, conjugate to the variable, $\delta$, is determined by,

$$q = \frac{\partial L}{\partial\dot{\delta}}, \tag{1.9}$$

and the Hamiltonian of the system can be obtained from the Legendre transform,

$$H(q,\delta) = q\dot{\delta} - L = \frac{q^2}{2m} + U(\delta). \tag{1.10}$$

The canonical form of the Hamiltonian is suitable for the transition to the quantum-mechanical description of the system. Considering the variables, $q$ and $\delta$, to be quantum-mechanical operators, satisfying the commutation relations,

$$[\delta, q] = i\hbar, \tag{1.11}$$

we transform the classical problem of the phase evolution into a quantum-mechanical one. In the $\delta$-representation, the operator, $q$, is given by $q = -i\hbar\partial/\partial\delta$. In what follows, we will use a similar scheme to derive circuit Hamiltonians for more complex systems.

The following step is to obtain the eigenfunctions and eigenvalues of the Hamiltonian (1.10). These depend entirely on the explicit form of the potential energy, $U(\delta)$. Fig. 2 illustrates a fragment of the dependence, $U(\delta)$. It can be easily seen that its shape resembles a tilted washboard.

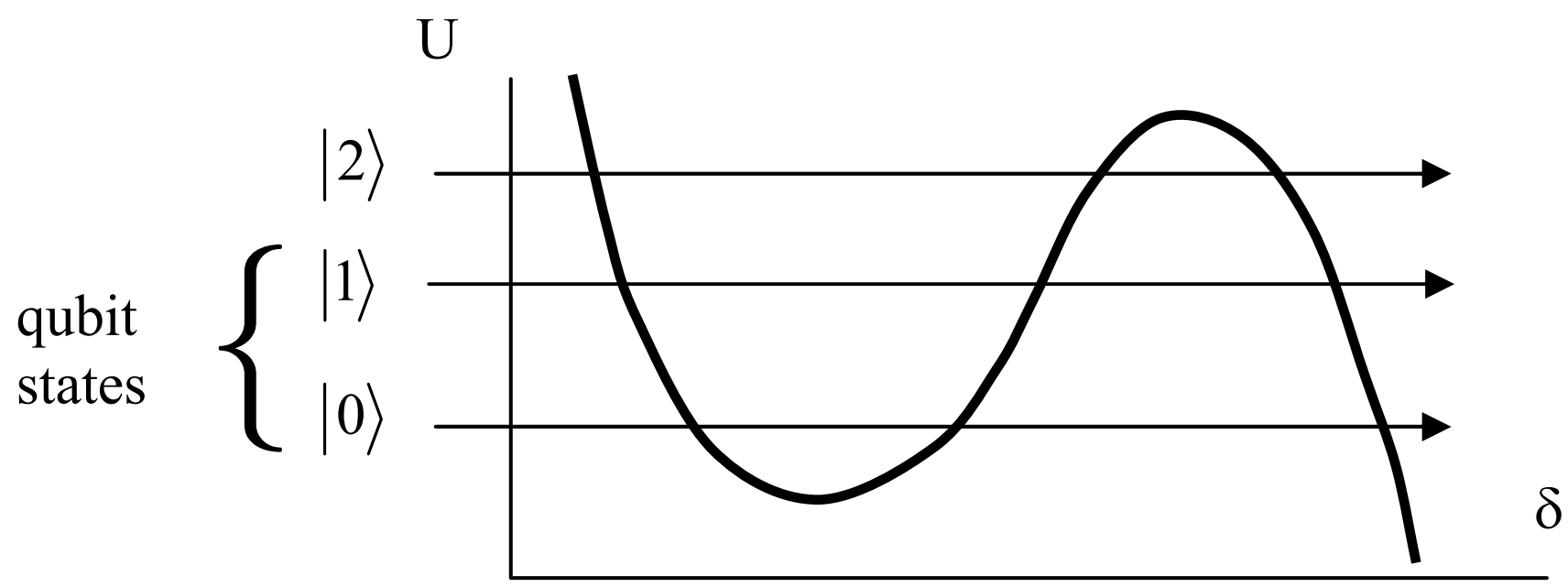

Fig. 2. Potential energy *vs* $\delta$. Three quasilocal levels in the minima are shown. The two lowest levels, $|0\rangle$ and $|1\rangle$, can be used as qubit states. The arrows indicate the possibility to tunnel to the neighboring minimum.

It is useful to think of this system as an anharmonic "LC" oscillator created from the Josephson inductance and the junction capacitance, with the anharmonicity arising from the nonlinear term, $1/\cos\delta$ in $L_e$. The lowest levels ($E_0, E_1$ and $E_2$) are not equidistant (as can be seen in Fig. 2). In this case the bias current at resonant frequency, $\Omega = (E_1 - E_0)/\hbar$, which is often used for controlling the qubit states, does not affect the level $|2\rangle$. To deal with only the two lowest levels in the course of qubit exploration, the temperature, $T$, is also considered to be sufficiently low to exclude significant thermal occupancy of the levels $|1\rangle$ and $|2\rangle$. The corresponding inequality,

$$\hbar\Omega \gg k_B T ,$$

is called a "quantum limit". A realization of these conditions is of great importance for the best superconducting qubit performance.

To readout the qubit state, an additional current pulse can be applied to the junction in order to lower the barrier height. The lowering should be sufficient for the transition from the state $|1\rangle$ but still small for the transition from the level $|0\rangle$. The transition can be easily registered experimentally since, in the case of tunneling, the junction exposes the ohmic resistance [3].

### 1.4. Flux-biased phase qubit

The rf-SQUID, which consists of a Josephson junction enclosed in a superconducting loop, can also be used as a qubit. A schematic of this situation is shown in Fig. 3.

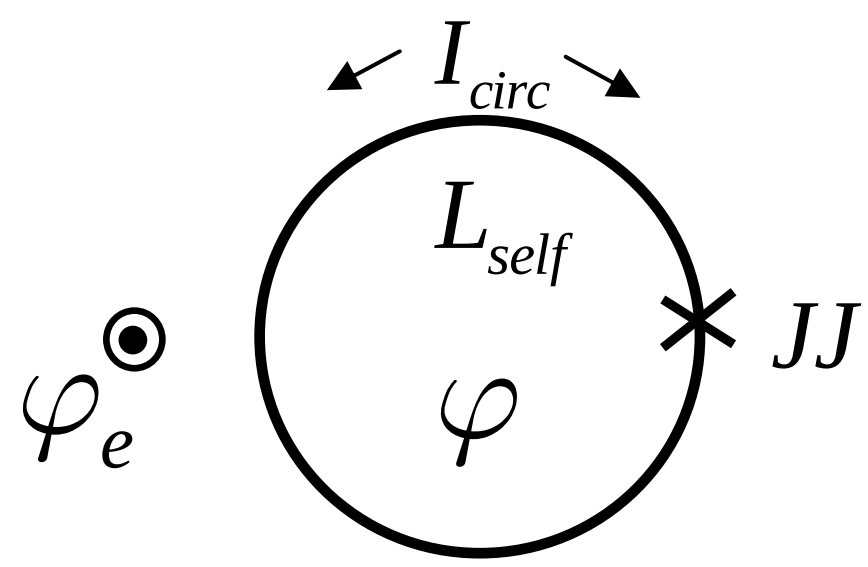

Fig. 3. A superconducting loop with a Josephson junction, which is exposed to the external flux, $\varphi_e$. A circulating current is generated which leads to a resulting flux, $\varphi$, inside the loop.

When an external magnetic field is applied to the loop, a screening current, $I_{circ}$, circulates in the ring. The screening current produces a magnetic flux, $L_{self} I_{circ}$, which changes the total flux. The total flux is

$$\varphi = \varphi_e - L_{self} I_{circ}, \tag{1.12}$$

where $L_{self}$ is the self-inductance of the loop. The magnetic energy of the loop is given by

$$\frac{1}{2} L_{self} (I_{circ})^2 = \frac{1}{2} \frac{(\varphi - \varphi_e)^2}{L_{self}}. \tag{1.13}$$

The current, $I_{circ}$, is determined by Eq. (1.1), where the phase, $\delta$, can be expressed via the total flux, $\varphi$. A simple explanation of the relationship between $\delta$ and $\varphi$ is given in Ref. 6. We will now shortly outline the most important points of that argumentation.

Let us consider the contact region in more detail (see Fig . 4). A superconductive electric current through the dielectric barrier is possible due to overlapping of the wave functions $\Psi_{1,2}(\theta)$, where $\theta = \theta(\vec{r})$ is the wave function phase ($\Psi_{1,2} \sim e^{i\theta}$). The quantum-mechanical operator describing the velocity of superconducting phase flow is given by,

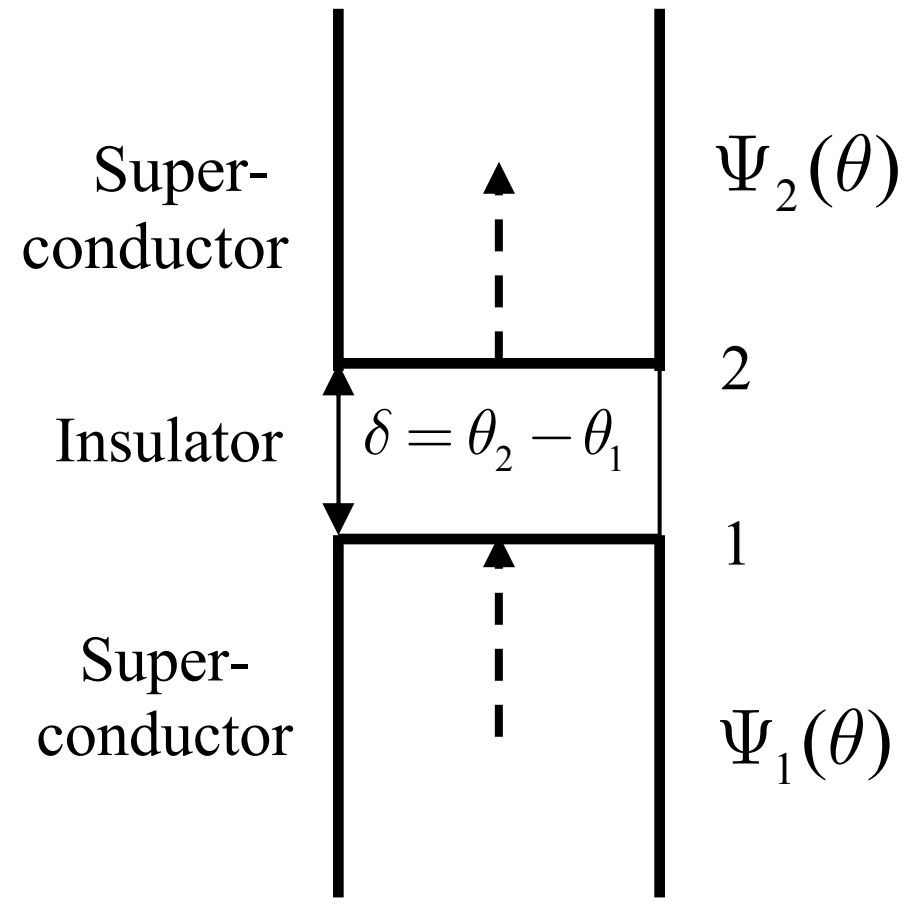

Fig. 4. Schematics of a Josephson junction. The dashed arrows indicate the integration path inside the body of the ring.

$$\vec{v} = \frac{1}{m_c}(-i\hbar\frac{\partial}{\partial\vec{r}} + 2e\vec{A}), \tag{1.14}$$

where, $m_c$, is the Cooper pair mass, $\vec{A}$ is the vector potential of the electromagnetic field. The action of $\vec{v}$ on the wave function results in

$$\vec{v}\Psi(\theta) = \frac{1}{m_c}(\hbar\vec{\nabla}\theta + 2e\vec{A})\Psi(\theta). \tag{1.15}$$

The quantity, $(\hbar\vec{\nabla}\theta + 2e\vec{A})/m_c$, is considered to be a velocity of a superconducting phase flow. Well inside the superconductor, for example along the arrows shown in Fig. 4, the current density is zero; so Eq. (1.15) gives

$$\hbar\vec{\nabla}\theta = -2e\vec{A}. \tag{1.16}$$

It is important to emphasize that Eq. (1.16) concerns only the regions deep inside the superconductors. It is not applicable for very thin conductors with the thickness less or of the order of the magnetic field penetration depth.

Now we will integrate both functions in Eq. (1.16) over the ring (as indicated in Fig. 4). The following relationships will be useful:

$$\oint d\vec{l}\vec{A}(\vec{r}) = \int d\vec{s}\vec{\nabla}\times\vec{A} = \int d\vec{s}\vec{H} = \varphi, \tag{1.17}$$

where $\vec{H}$ is the magnetic field which defines the flux, $\varphi$ through the loop, $d\vec{s}$ is an element of the surface confined by the ring; the vector, $d\vec{s}$, is oriented perpendicular to this surface element.

Using Eq. (1.17), we get after integration,

$$\delta = \theta_2 - \theta_1 = 2\pi\varphi / \varphi_0 . \tag{1.18}$$

Then, the Hamiltonian of the system corresponding to Fig. 3 is given by

$$H = \frac{q^2}{2m} - E_J \cos\left(2\pi \frac{\varphi}{\varphi_0}\right) + \frac{(\varphi - \varphi_e)^2}{2L_{self}} . \tag{1.19}$$

The two lowest eigenstates of the Hamiltonian (1.19), $|0\rangle$ and $|1\rangle$, can be chosen as the qubit states. The Hamiltonian (1.19) depends on the external flux, $\varphi_e$, which can be the controlling parameter for the qubit states. There are specific values of $\varphi_e$ $[(\varphi_e / \varphi_0) = (n + 1/2), n = 0, \pm 1, \pm 2, ...]$, where the potential energy, $U$ ,

$$U = U(\varphi) = -E_J \cos\left(2\pi \frac{\varphi}{\varphi_0}\right) + \frac{(\varphi - \varphi_e)^2}{2L_{eff}} , \tag{1.20}$$

has two symmetric minima as shown in Fig. 5.

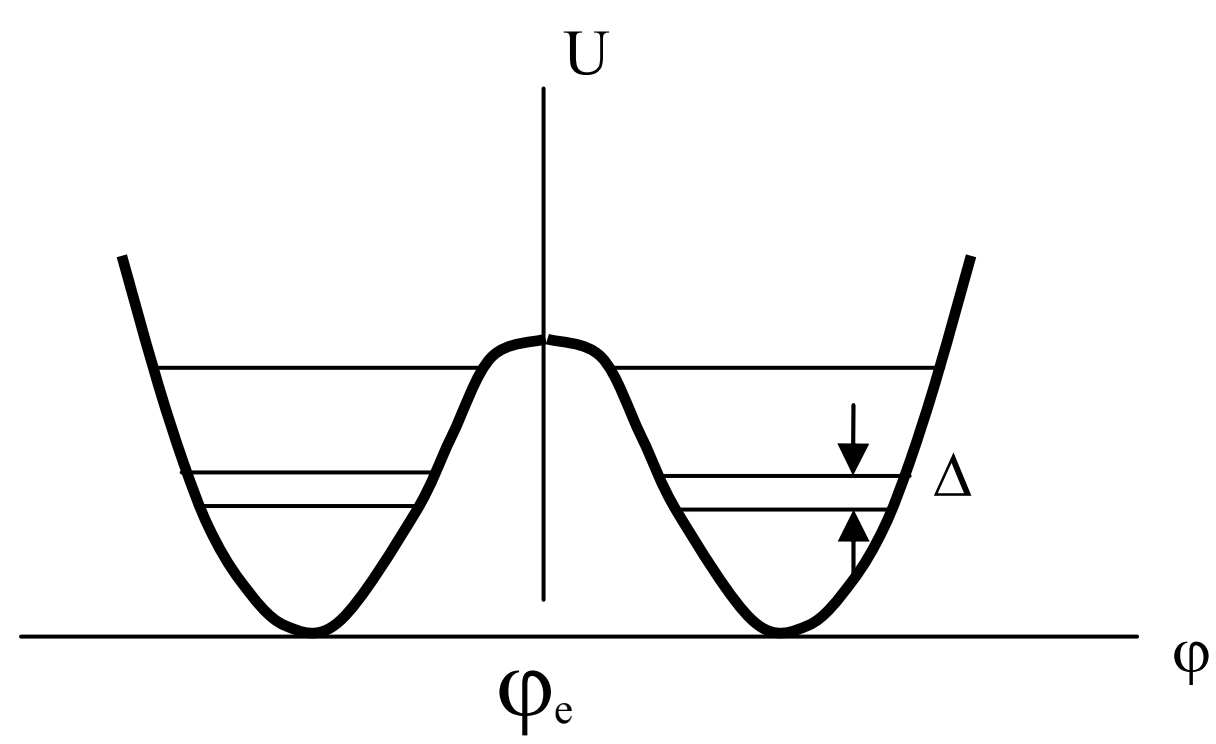

Fig. 5. Double-well potential of the rf-SQUID with degenerate quantum levels in the individual wells. The quantity, Δ , is the level splitting due to the macroscopic quantum tunneling (MQT).

If the tunneling through the barrier from the lowest level has low probability, then the level splitting is small and can be easily varied by means of variation of the external flux, $\varphi_e$. In the case that the tunneling barrier is much smaller than the Josephson energy, $E_J$, the potential energy in the vicinity of $\varphi_e = \varphi_0 / 2$ can be approximated as [3]

$$U \approx E_L\left[1+\chi-\frac{\chi}{2}\tilde{\delta}^2+(1+\chi)\frac{\tilde{\delta}^4}{24}-\tilde{\delta}f\right], \tag{1.21}$$

where

$$E_L=\frac{(\varphi_0/2\pi)^2}{L_{self}},\ \chi=\frac{E_J}{E_L}-1,\ 0<\chi<<1,\ \tilde{\delta}=\pi\left(\frac{2\varphi}{\varphi_0}-1\right),\ f=\pi\left(\frac{2\varphi_e}{\varphi_0}-1\right).$$

The central maximum in Fig. 5 is higher than the minimum values by $\frac{3}{2}E_L\chi^2$. The parameter, $f$, determines the asymmetry of the potential energy which arises when $\varphi_e$ is slightly different from $\varphi_0/2$. For theoretical analysis of the qubit states dynamics, the formalism of two-level tunneling states [7], developed for the case of glasses, is widely used in the literature. Considering the tunneling probability to be small (but still finite), the potential shown in Fig. 5 is approximated by a double-well potential with $U_l$ and $U_r$ as shown in Fig. 6. The ground state eigenfunctions in the isolated wells are supposed to be $|l\rangle$ and $|r\rangle$,

$$H_l|l\rangle=E_l|l\rangle,\quad H_r|r\rangle=E_r|r\rangle. \tag{1.22}$$

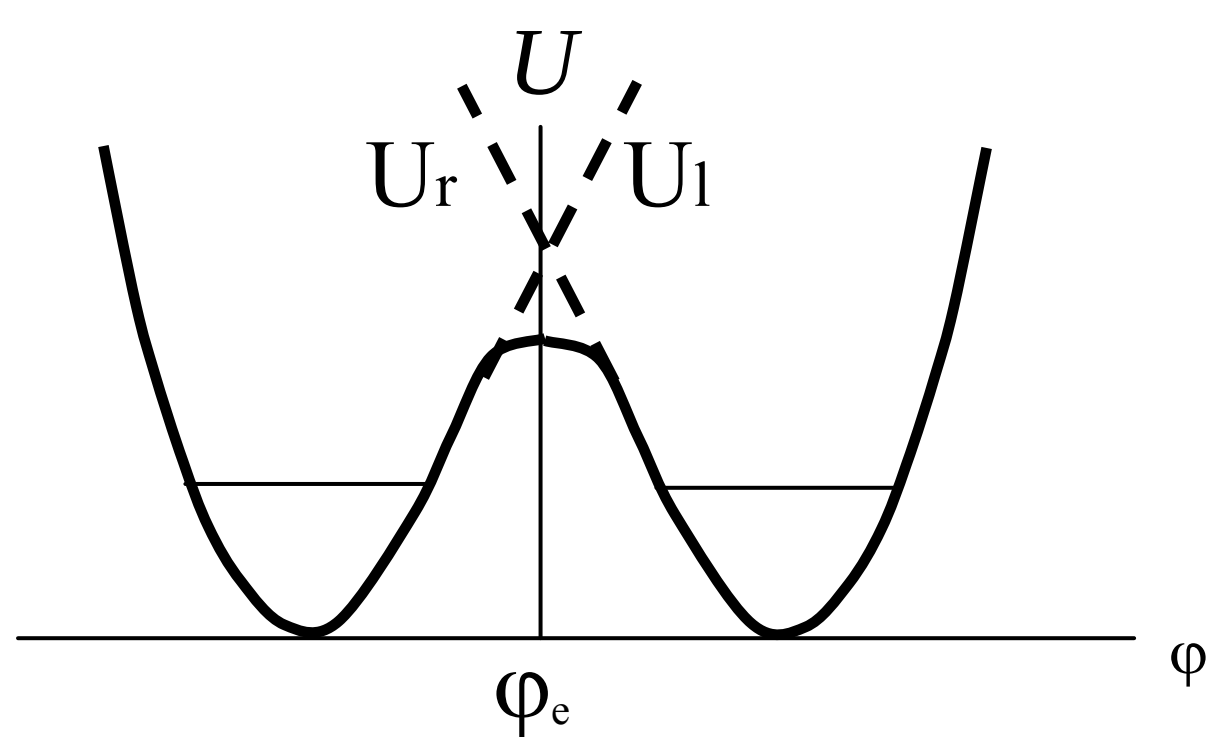

Fig. 6. Dashed lines indicate potentials of the left and right wells when $f=0$.

The complete Hamiltonian, *H,* can be written as,

$$H=H_l+(U-U_l)=H_r+(U-U_r). \tag{1.23}$$

Considering the states $|l\rangle$ and $|r\rangle$ as the basis, we can write the Hamiltonian matrix in the local representation (in $|l\rangle,|r\rangle$ basis) as

$$H = \begin{vmatrix} E_l + \langle l|U - U_l|l\rangle & \langle l|H|r\rangle \\ \langle r|H|l\rangle & E_r + \langle r|U - U_r|r\rangle \end{vmatrix}. \tag{1.24}$$

If the extension of each localized wave function into the barrier is small, the terms with $U - U_{l,r}$ can be neglected in comparison with $E_{l,r}$, and if the zero of energy is chosen as the mean of $E_l$ and $E_r$, the Hamiltonian in Eq. (1.24) becomes

$$H = -\frac{1}{2}\begin{vmatrix} \varepsilon & \Delta \\ \Delta & -\varepsilon \end{vmatrix}, \tag{1.25}$$

where $\varepsilon = E_r - E_l$, $\Delta = -2\langle l|H|r\rangle$, $\langle l|H|r\rangle = \langle r|H|l\rangle$. The Hamiltonian matrix (1.25) can be rewritten in terms of Pauli spin matrices as

$$H = -\frac{1}{2}(\varepsilon\sigma_z + \Delta\sigma_x). \tag{1.26}$$

The eigenvalues of the Hamiltonian (1.26) are

$$E_{1,2} = \mp\frac{1}{2}\sqrt{\varepsilon^2 + \Delta^2}, \tag{1.27}$$

and the eigenfunctions can be written in the form

$$\Psi_{1,2} = \frac{\left(\sqrt{\varepsilon^2 + \Delta^2} \pm \varepsilon\right)^{1/2}}{\sqrt{2}\left(\varepsilon^2 + \Delta^2\right)^{1/4}}\left(|l\rangle - \frac{\varepsilon \mp \sqrt{\varepsilon^2 + \Delta^2}}{\Delta}|r\rangle\right). \tag{1.28}$$

For positive functions, $|l\rangle$ and $|r\rangle$, the quantity $\Delta$ is also positive as can be seen from the following relations:

$$\Delta = -2\langle l|H|r\rangle = -2\langle l|H_r + (U - U_r)|r\rangle \approx -2\langle l|(U - U_r)|r\rangle,$$

where $U - U_r < 0$ (see Fig. 6). Therefore, the function $\Psi_1(\varphi)$ does not change sign at any $\varphi$, as it should be in the case of the lowest level. This conclusion remains unchanged for the case when the functions $|l\rangle$ and $|r\rangle$ are chosen with different signs.

In the diagonal representation, where the functions $\Psi_{1,2}(\varphi)$ form the basis, the Hamiltonian is given by

$$H = -\frac{\sqrt{\varepsilon^2 + \Delta^2}}{2}\sigma_z. \tag{1.29}$$

### 1.5. Two phase qubits coupling with a resonant cavity

We have considered the simplest realizations of individual qubits which states can be affected by the external forces such as an electric current or magnetic flux. The quantum computing requires joint work of many interconnected qubits. The coherent transfer of quantum states between qubits is one of the most important problems in quantum computation. The transfer can be realized through a quantum bus. This quantum bus can be a resonant superconducting transmission line [8].

Now we consider the Hamiltonian of two qubits connected by a resonance line. Its electrical scheme is shown in Fig. 7. External controlling fields affecting the *JJ* junctions (see Figs. 1,3) are not shown. A mathematical description of this system should be based on quantum-mechanical approach. The Hamiltonian description of the dynamics of electrical circuits is given in details in Ref. [9]. The simplest situation is in the absence of the dissipative elements. Formally, any circuit can be considered as a network whose branches consist of two-terminal elements. The element of each branch, $b$, is characterized by two variables: the voltage, $v_b(t)$, across it and the current, $i_b(t)$, flowing through it (see Fig. 8). The Hamiltonian description requires introduction of branch fluxes and branch charges, which are defined by $v_b(t)$ and $i_b(t)$ as

$$\varphi_b(t) = \int_{-\infty}^{t} v_b(t')dt', \quad q_b(t) = \int_{-\infty}^{t} i_b(t')dt' . \tag{1.30}$$

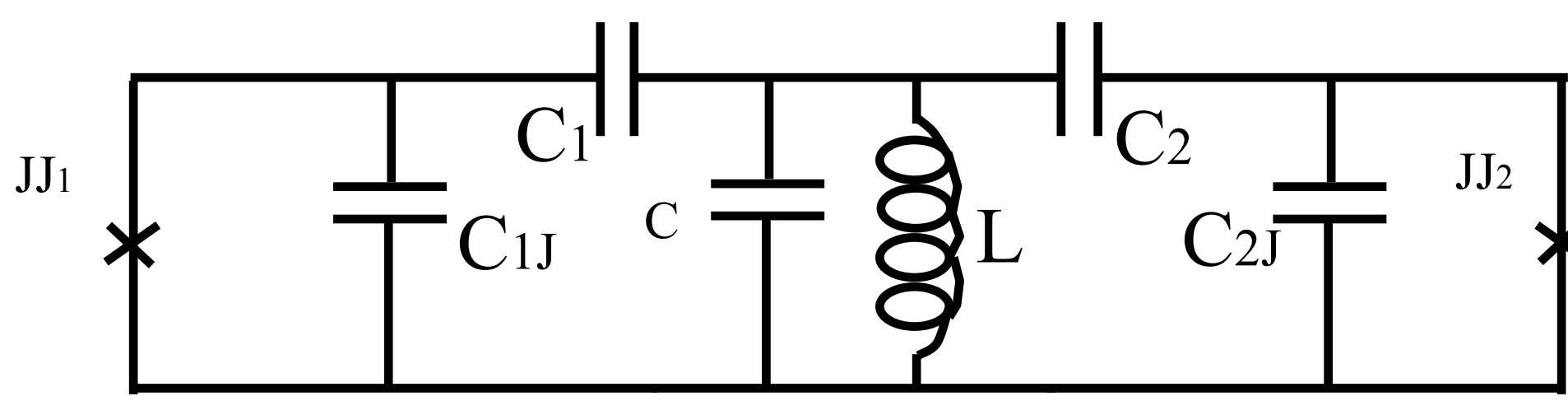

Fig. 7. The Josephson junctions, $JJ_1$ and $JJ_2$ , coupled to the resonator which is formed by the capacitor, C, and inductance, L. $C_{1J}$ and $C_{2J}$ are the capacitors of the JJ junctions, $C_{1,2}$ – coupling capacitors.

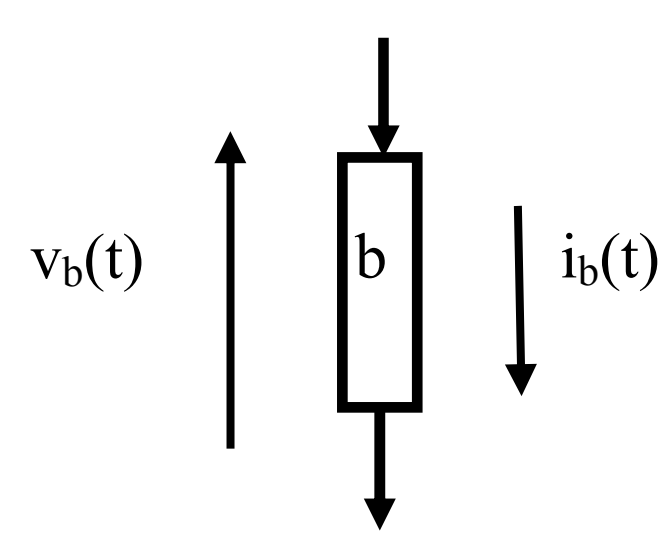

Fig. 8. Two-terminal element characterized by branch variables, $v_b(t)$ and $i_b(t)$.

This choice of the variable $\varphi_b(t)$ is convenient for the description of the Josephson contact because the phase, $\delta$, (see Eq. (1.1)) is equal to the value of the corresponding flux times $2\pi/\varphi_0$ (see Eq. (1.2)). It can be easily seen from the definition of the variable, $\varphi_b(t)$, that the current through any capacitor, $C$, is given by $i_C = C\ddot{\varphi}$. Similarly, the current through an inductance, $L$, is given by $\frac{1}{L}\varphi$. The "kinetic energy" in the circuit shown in Fig. 7 is given by

$$T = \frac{C_{1J}}{2}\dot{\varphi}_{1J}^2 + \frac{C_1}{2}(\dot{\varphi} - \dot{\varphi}_{1J})^2 + \frac{C}{2}\dot{\varphi}^2 + \frac{C_2}{2}(\dot{\varphi} - \dot{\varphi}_{2J})^2 + \frac{C_{2J}}{2}\dot{\varphi}_{2J}^2, \tag{1.31}$$

where we have used Eq. (1.30) and Kirchhoff's voltage law which give

$$\dot{\varphi}_1 = \dot{\varphi}_C - \dot{\varphi}_{1J}, \quad \dot{\varphi}_2 = \dot{\varphi}_{2J} - \dot{\varphi}_C. \tag{1.32}$$

The indices in voltage variables, $\varphi_\alpha$, correspond to those of capacitors.

The "potential" energy is given by

$$U = -E_{1J}\cos\left(2\pi\frac{\varphi_{1J}}{\varphi_0}\right) - E_{2J}\cos\left(2\pi\frac{\varphi_{2J}}{\varphi_0}\right) + \frac{\varphi^2}{2L}. \tag{1.33}$$

The circuit Lagrangian, $L = T - U$, is expressed in terms of the flux variables, and similar to Eq. (1.9) we can obtain the canonical conjugate momenta, $q_i$, from

$$q_i = \frac{\partial L}{\partial \dot{\varphi}_i}. \tag{1.34}$$

After solution of the system of three linear equations for the momenta, $q_i$, we can obtain the Hamiltonian which, in the case of "symmetric" parameters $E_{1J} = E_{2J} \equiv E_J$, $C_{1J} = C_{2J} \equiv C_J$, $C_1 = C_2 \equiv C_c$, and small coupling capacitor, $C_c$, is given by

$$H = \sum_i q_i\dot{\varphi}_i - L \approx \frac{q_{1J}^2 + q_{2J}^2}{2C_J}\left(1 - \frac{C_c}{C_J}\right) + \frac{q^2}{2C}\left(1 - \frac{2C_c}{C}\right) + q\left(q_{1J} + q_{2J}\right)\frac{C_c}{C_J C} -$$

$$E_J\left[\cos\left(\frac{2\pi}{\varphi_0}\varphi_{1J}\right) + \cos\left(\frac{2\pi}{\varphi_0}\varphi_{2J}\right)\right] + \frac{\varphi^2}{2L}. \tag{1.35}$$

As previously, the transition to a quantum-mechanical description can be undertaken by stating the commutation relations between conjugate variables, $q_i$ and $\varphi_i$. Then the dynamics of the system is governed by the Schrödinger equation

$$i\hbar\partial_t\Psi = H(q_i,\varphi_i)\Psi\,, \tag{1.36}$$

where $q_i = -i\hbar\dfrac{\partial}{\partial\varphi_i}$, $\Psi \equiv \Psi(\varphi,\varphi_{1J},\varphi_{2J})$.

## 1.6. A coupling of qubit and resonator states

The interaction of qubit and resonator sub-systems imposes significant correlations between their states that can be used for nondestructive measurements of one of these. Measuring the state of one of the sub-systems provides the information about the other one (see, for example, Refs. 10 and 11). In what follows, we analyze this phenomenon in more detail. As previously, we consider the case of weak qubit-resonator coupling. The equivalent circuit is shown in Fig. 9.

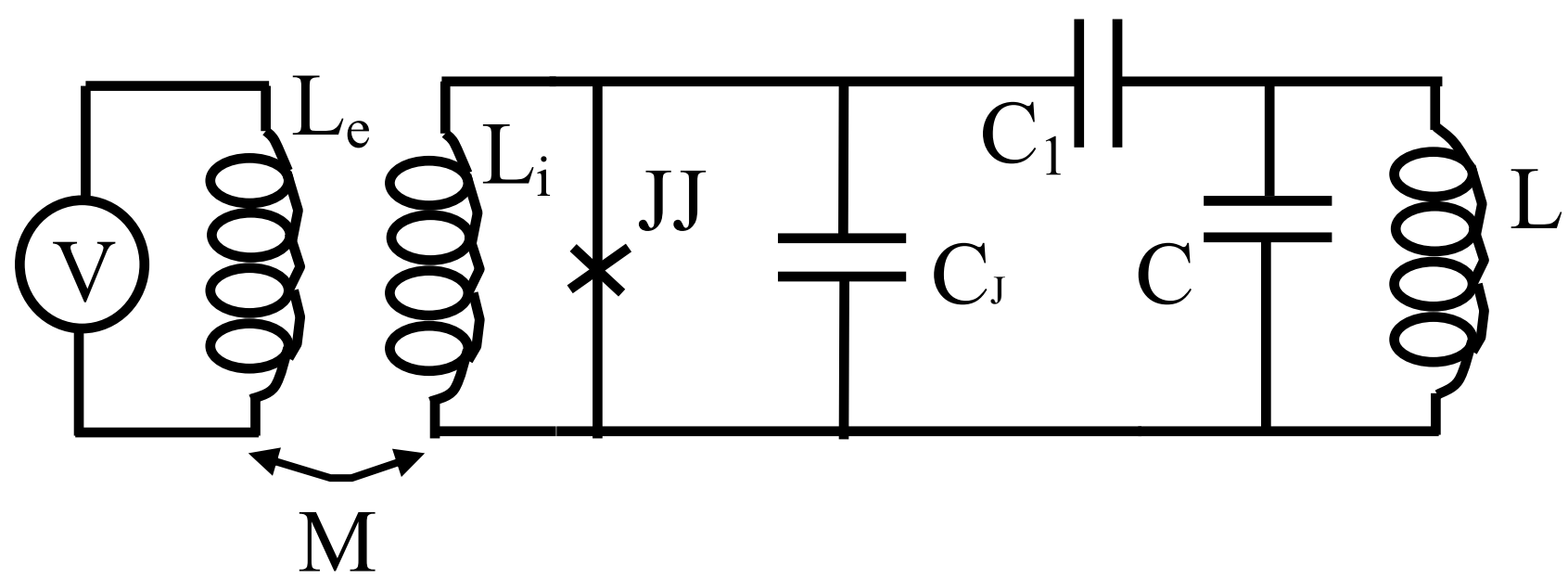

Fig. 9. An external voltage, $V$, is a source of the biased magnetic flux in the qubit circuit. $M$ is the mutual inductance between coils, $L_e$ and $L_i$. The effect of the input-coil ($L_i$) current on the external-coil ($L_e$) current is neglected.

The external voltage, $V$, provides an external magnetic flux through the mutual inductance, $M$, into the rf-SQUID loop thus ensuring the most favorable choice of its operation. The total Hamiltonian of the circuit can be derived as was explained before. It is given by

$$H = \frac{q_J^2}{2C_J}\left(1-\frac{C_1}{C_J}\right) + \frac{q^2}{2C}\left(1-\frac{C_1}{C}\right) - E_J\cos\left(\frac{2\pi}{\varphi_0}\varphi_J\right) + \frac{\left(\varphi_J-\varphi_e\right)^2}{2L_i} + \frac{\varphi^2}{2L} + C_1\frac{qq_J}{CC_J}\,. \tag{1.37}$$

The qubit system, described by the Hamiltonian

$$H_q = \frac{q_J^2}{2C_J}\left(1-\frac{C_1}{C_J}\right) - E_J\cos\left(\frac{2\pi}{\varphi_0}\varphi_J\right) + \frac{\left(\varphi_J-\varphi_e\right)^2}{2L_i}, \tag{1.38}$$

and the resonator with the Hamiltonian,

$$H_r = \frac{q^2}{2C}\left(1-\frac{C_1}{C}\right) + \frac{\varphi^2}{2L}, \tag{1.39}$$

are coupled via the capacitor, $C_1$. The corresponding interaction term is given by,

$$H_{\text{int}} = C_1\frac{qq_J}{CC_J}. \tag{1.40}$$

As we see, both Hamiltonians, $H_q$ and $H_r$, depend on the capacitor, $C_1$, that is the effect of qubit-resonator interaction. Also the resonator quantum state depends on the qubit state and vice versa. We will see this dependence in an explicit form below. As previously, we restrict our analysis by only two lowest states (the qubit states) of the Hamiltonian (1.38). Then Eq. (1.38) in the diagonal representation is given by Eq. (1.29)

$$H_q = -\sqrt{\varepsilon^2+\Delta^2}\sigma_z \equiv -\frac{\hbar\omega_q}{2}\sigma_z. \tag{1.41}$$

We consider here that an adequate choice of the external flux and parameters of the JJ junction provides a double-well potential similar to that shown in Fig. 5.

The resonator Hamiltonian can be rewritten as

$$H_r = \hbar\omega_r\left(a^+a+\frac{1}{2}\right), \tag{1.42}$$

where $\omega_r^2 = \frac{1}{L\left(C+C_1\right)}$, $a^+, a$ are the creation and annihilation operators of the resonator excitations.

Some explanations are required while expressing $H_{\text{int}}$ in terms of $a^+$, $a$ and Pauli matrices. The diagonal matrix elements of $q_J$ in the basis of $\Psi_{1,2}(\varphi_J)$ (see Eq. (1.28)) are equal to zero because no tunneling from the double-well region is supposed. The non-diagonal terms are given by,

$$\left\langle\Psi_1\left|q_J\right|\Psi_2\right\rangle = -\left\langle\Psi_2\left|q_J\right|\Psi_1\right\rangle = i\hbar\left\langle l\right|\frac{\partial}{\partial\varphi_J}\left|r\right\rangle. \tag{1.43}$$

Therefore the operator $q_J$ can be written in the $\Psi_{1,2}$ basis as

$$q_J = -\hbar\left\langle l\right|\frac{\partial}{\partial\varphi_J}\left|r\right\rangle\sigma_y. \tag{1.44}$$

The oscillator momentum, $q$, is given by

$$q=-i\left(\frac{\hbar}{2}(C+C_1)\omega_r\right)^{1/2}\left(a^+-a\right)\equiv-i\left(\frac{\hbar}{2}\sqrt{\frac{C+C_1}{L}}\right)^{1/2}\left(a^+-a\right). \tag{1.45}$$

From Eqs. (1.44) and (1.45) we get

$$H_{\text{int}}=i\lambda\sigma_y\left(a^+-a\right), \tag{1.46}$$

where

$$\lambda=\hbar^{3/2}\frac{C_1}{C_J}\left(\frac{\omega_r}{2C}\right)^{1/2}\langle l|\frac{\partial}{\partial\varphi_J}|r\rangle .$$

The energy spectrum of a non-interacting qubit-resonator system is given by

$$E_{n\uparrow}^{(0)}=-\frac{\hbar\omega_q}{2}+\hbar\omega_r\left(n+\frac{1}{2}\right),\ E_{n\downarrow}^{(0)}=\frac{\hbar\omega_q}{2}+\hbar\omega_r\left(n+\frac{1}{2}\right). \tag{1.47}$$

The notation, $\uparrow/\downarrow$, denotes the qubit in the ground/excited state, $n$ indicates the number of photons in the oscillator. The spectrum is disturbed by the interaction Hamiltonian. The second-order perturbation expansion gives the displacements of levels:

$$E_{n\uparrow}^{(2)}=-\sum_{m\neq n}\frac{\left|\langle m\downarrow|H_{\text{int}}|n\uparrow\rangle\right|^2}{E_{m\downarrow}-E_{n\uparrow}},\quad E_{n\downarrow}^{(2)}=-\sum_{m\neq n}\frac{\left|\langle m\uparrow|H_{\text{int}}|n\downarrow\rangle\right|^2}{E_{m\uparrow}-E_{n\downarrow}}. \tag{1.48}$$

A straightforward calculation results in (see also [5])

$$E_{n\uparrow}^{(2)}=-\lambda^2\left[\frac{n+1}{\hbar\left(\omega_q+\omega_r\right)}+\frac{n}{\hbar\left(\omega_q-\omega_r\right)}\right],\quad E_{n\downarrow}^{(2)}=\lambda^2\left[\frac{n+1}{\hbar\left(\omega_q-\omega_r\right)}+\frac{n}{\hbar\left(\omega_q+\omega_r\right)}\right]. \tag{1.49}$$

As we see, in the case $\omega_q>\omega_r$, the shift of oscillator levels is negative for the ground state of the qubit and positive for the excited state. Eqs. (1.49) can be also treated as the shift of qubit levels due to the influence of the resonator. In the specific case of $n=0$ (the vacuum state of the resonator), the qubit frequency variation is given by

$$\frac{2\lambda^2}{\hbar^2}\frac{\omega_q}{\omega_q^2-\omega_r^2}, \tag{1.50}$$

which can be regarded as the effect of the Lamb shift of qubit levels. (The qubit system interacts with the vacuum state of the resonator.)

Eqs. (1.49) can be used for experimental determination of the qubit state by measuring the oscillator frequency shift, or for determining the number of photons in the resonator by measuring the resonant frequency of the qubit. Note that for a given state of the qubit the energy shift is the same for all the oscillator states. As a result, the frequency of the allowed transitions (between the neighboring energy levels) remains independent of the energy level, *n*. With accuracy to the order of the terms proportional to $\lambda^2$, the effective Hamiltonian of the system "qubit+resonator" can be written as

$$\begin{aligned} \mathrm{H} &= -\frac{\hbar\omega_q}{2}\sigma_z + \hbar\left(\omega_r + \delta\omega_r\sigma_z\right)\left(a^\dagger a + \frac{1}{2}\right), \\ \delta\omega_r &= -\frac{2\lambda^2}{\hbar^2}\frac{\omega_q}{\omega_q^2 - \omega_r^2}. \end{aligned} \tag{1.51}$$

In this approximation, the effective interaction Hamiltonian

$$\hbar\delta\omega_r\sigma_z\left(a^\dagger a + \frac{1}{2}\right) \tag{1.52}$$

commutes with the Hamiltonian of the qubit. This means that the interaction between the resonator and qubit allows one to implement a nondestructive measurement of a qubit state.

## 1.7. The nonlinear resonator

The situation changes if we take into consideration higher orders of the perturbation theory. The forth order correction to the energy levels makes energy levels non-equidistant. For the Hamiltonians (1.41), (1.42), (1.46) we have, up to the fourth order by the perturbation parameter, $\lambda$,

$$E_{n\uparrow} = E_{n\uparrow}^{(0)} + E_{n\uparrow}^{(2)} + E_{n\uparrow}^{(4)},\ E_{n\downarrow} = E_{n\downarrow}^{(0)} + E_{n\downarrow}^{(2)} + E_{n\downarrow}^{(4)}, \tag{1.53}$$

where $E_{n\uparrow,\downarrow}^{(0)}$ and $E_{n\uparrow,\downarrow}^{(2)}$ are given in (1.47)-(1.49), and the forth order corrections which are responsible for nonlinear terms are

$$\begin{aligned} E_{n\uparrow}^{(4)} &= \frac{\lambda^4}{\hbar^3}\frac{\left[\left(\omega_q-\omega_r\right)(n+1)+\left(\omega_q+\omega_r\right)n\right]\left[\left(\omega_q-\omega_r\right)^2(n+1)+\left(\omega_q+\omega_r\right)^2 n\right]}{\left(\omega_q^2-\omega_r^2\right)^3} + \\ &\frac{\lambda^4}{2\omega_r\hbar^3}\frac{\left[-\left(\omega_q-\omega_r\right)^2(n+1)(n+2)+\left(\omega_q+\omega_r\right)^2 n(n-1)\right]}{\left(\omega_q^2-\omega_r^2\right)^2}, \end{aligned} \tag{1.54}$$

$$E_{n\downarrow}^{(4)}=-\frac{\lambda^4}{\hbar^3}\frac{\left[\left(\omega_q+\omega_r\right)\left(n+1\right)+\left(\omega_q-\omega_r\right)n\right]\left[\left(\omega_q+\omega_r\right)^2\left(n+1\right)+\left(\omega_q-\omega_r\right)^2 n\right]}{\left(\omega_q^2-\omega_r^2\right)^3}+ \frac{\lambda^4}{2\omega_r\hbar^3}\frac{\left[-\left(\omega_q+\omega_r\right)^2\left(n+1\right)\left(n+2\right)+\left(\omega_q-\omega_r\right)^2 n\left(n-1\right)\right]}{\left(\omega_q^2-\omega_r^2\right)^2}. \tag{1.55}$$

For conditions close to the resonance, $\omega_q \sim \omega_r$, but still valid for the perturbative approach, the energy shifts of the resonator proportional to $\lambda^4$ can be written as

$$E_{n\uparrow}^{(4)}=\frac{\lambda^4\left(\omega_q+\omega_r\right)^3 n^2}{\hbar^3\left(\omega_q^2-\omega_r^2\right)^3},\ E_{n\downarrow}^{(4)}=-\frac{\lambda^4\left(\omega_q+\omega_r\right)^3\left(n+1\right)^2}{\hbar^3\left(\omega_q^2-\omega_r^2\right)^3}. \tag{1.56}$$

For the non-resonant case, $\omega_q \gg \omega_r$, the corresponding energy shifts are

$$E_{n\uparrow}^{(4)}=\frac{6\lambda^4 n^2}{\hbar^3\omega_q^3},\ E_{n\downarrow}^{(4)}=-\frac{6\lambda^4 n^2}{\hbar^3\omega_q^3}. \tag{1.57}$$

It follows from (1.47), (1.49), (1.53)-(1.55) that up to the fourth order in the perturbation parameter, $\lambda$, the effective Hamiltonian of the resonator can be represented in the form of a quantum nonlinear oscillator,

$$H_{\mp}=\hbar\omega_{r,\mp}a^\dagger a+\mu_{\mp}\hbar^2\left(a^\dagger a\right)^2+\varsigma_{\mp}, \tag{1.58}$$

where $\mp$ denotes the state of a qubit, ↑ or ↓, correspondingly, $\omega_{r,\mp}$ is a renormalized frequency, $\mu_{\mp}$ is a parameter of nonlinearity, and $\varsigma_{\mp}$ is a constant.

It is important to note that if initially the resonator is populated in the coherent state,

$$\left|\alpha\right\rangle=e^{-|\alpha|^2/2}\sum_{n=0}^{\infty}\frac{\alpha^n}{\sqrt{n!}}\left|n\right\rangle,\ \left(a^\dagger a\left|n\right\rangle=n\left|n\right\rangle\right), \tag{1.59}$$

then, after some time, $t_\hbar$, the coherence of the resonator field will be lost (see [12] and references therein). For example, it is easy to show that the dynamics of the average value of the operator, $a$, in the initially coherent state (1.59) is given by the expression

$$\alpha\left(\tau\right)\equiv\left\langle\alpha\left|a\left(\tau\right)\right|\alpha\right\rangle=\alpha e^{-i(1+\bar{\mu})\tau}\exp\left\{\left|\alpha\right|^2\left[e^{-i2\bar{\mu}\tau}-1\right]\right\}, \tag{1.60}$$

where the dimensionless time and parameters were introduced,

$$\tau = \omega_r t,\ \bar{\mu} = \frac{\hbar\mu}{\omega_r}. \tag{1.61}$$

In (1.60), (1.61) and below we omitted the index "$\mp$", and assumed that $\omega_r$ is a renormalized frequency of a resonator, $\omega_{r,-}$ or $\omega_{r,+}$.

The solution (1.60) has three characteristic time-scales [12]. We consider the quasiclassical case for the resonator field when the average number of photons is large enough, $\bar{n} = |\alpha|^2 \gg 1$. In the limit, $\bar{\mu}\tau \ll 1$, (1.60) can be represented in the form

$$\begin{aligned} &\alpha(\tau) = \alpha_{cl}(\tau) e^{-\tau^2/2\tau_\hbar^2}\left[1 + O(\bar{\mu}\tau) + O\left(|\alpha|^2 \bar{\mu}^3\tau^3\right)\right], \\ &\alpha_{cl}(\tau) = \alpha e^{-i(1+2\mu_{cl})\tau}, \end{aligned} \tag{1.62}$$

where $\alpha_{cl}(\tau)$ describes the classical dynamics, $\mu_{cl} = \mu J/\omega_r$ is the classical parameter of nonlinearity, and $J = \hbar|\alpha|^2$ is the classical action. The first time-scale is the characteristic "classical" time-scale, which can be chosen as a period of the classical nonlinear oscillations,

$$\tau_{cl} = \frac{2\pi\omega_r}{\omega_{cl}} = \frac{2\pi}{1+2\mu_{cl}}.$$

The second time-scale characterizes the departure of quantum dynamics from the corresponding classical one (see details in [12]),

$$\tau_\hbar = \frac{1}{2\bar{\mu}\sqrt{\bar{n}}}. \tag{1.63}$$

Finally, the third time-scale,

$$\tau_R = \frac{\pi}{\bar{\mu}}, \tag{1.64}$$

characterizes the time of quantum revivals, which in the quasiclassical region of parameters is large enough.

The frequency spectrum of $\alpha(\tau)$ (in the units $\omega_r$) in (1.60) consists of one central line with $\nu = \omega_{cl}/\omega_r = 1 + 2\mu_{cl}$, which characterizes the classical coherent oscillations, and a width (or envelope) related to the expression (1.62), which is approximately equal to

$$\Delta \nu_{\hbar} \approx \frac{2\sqrt{2}}{\tau_{\hbar}}. \tag{1.65}$$

The envelope of the frequency spectrum has an internal structure related to the frequencies of quantum revivals, $\nu_{R,n} = n\bar{\mu}/\pi$ [12].

We can envision two effects associated with the non-equidistance of the energy levels of the resonator. If one uses the quasiclassical superposition of the resonator states (such as a coherent state), the nonlinear corrections to the energy levels may cause a broadening of the resonator frequency given by (1.65) with the characteristic time-scale (1.63). This effect may be important for high quality resonators. In this case we can introduce a "quantum" quality factor of the resonator

$$Q_{\hbar} = \frac{1}{\Delta \nu_{\hbar}}. \tag{1.66}$$

Then, if the "standard" quality factor of the resonator, $Q$, exceeds the quantum quality factor, $Q_{\hbar}$ $\left(Q > Q_{\hbar}\right)$, the width of the frequency spectrum of the resonator will be defined not by the expression: $\Delta \nu_r = \nu_r / Q$, but by the quantum parameter, $\Delta \nu_{\hbar}$, which is presented in (1.65) in the dimensionless form. In this quasiclassical regime, the quantum corrections represent a singular perturbation to the classical dynamics. (See details in [12], and in references therein.)

On the other hand, if one exploits only two lowest levels of the resonator in order to measure the qubit states, then the non-equidistance of the energy levels could help to achieve this goal. In this case, the parameter of nonlinearity, $\bar{\mu}$, in (1.61) should be large enough.

### 1.8. Qubit-resonator inductive coupling

A circuit with the inductive qubit-resonator coupling is shown in Fig. 10. An external flux is supplied through the mutual inductance, $M_J$. Also an external source of *ac*-voltage is shown which varies the qubit state that can be registered after amplifying the alternating signal used for the qubit readout.

Equations of motion for this scheme can be derived as previously. For example, the kinetic energy is given by

$$T = \frac{C_V}{2}\left(\dot{\varphi}_J - V\right)^2 + \frac{C_J}{2}\dot{\varphi}_J^2 + \frac{C}{2}\dot{\varphi}^2, \tag{1.67}$$

where the Kirchhoff's voltage law was taken into account. The potential energy, $U$, is given by

$$U = -E_J \cos\left(\frac{2\pi}{\varphi_0}\varphi_J\right) + \frac{(\varphi_J - \varphi_e)^2}{2L_J} + \frac{\varphi_J^2}{2L_1} + \frac{\varphi^2}{2L} - \frac{M}{LL_1}\varphi\varphi_J\,. \tag{1.68}$$

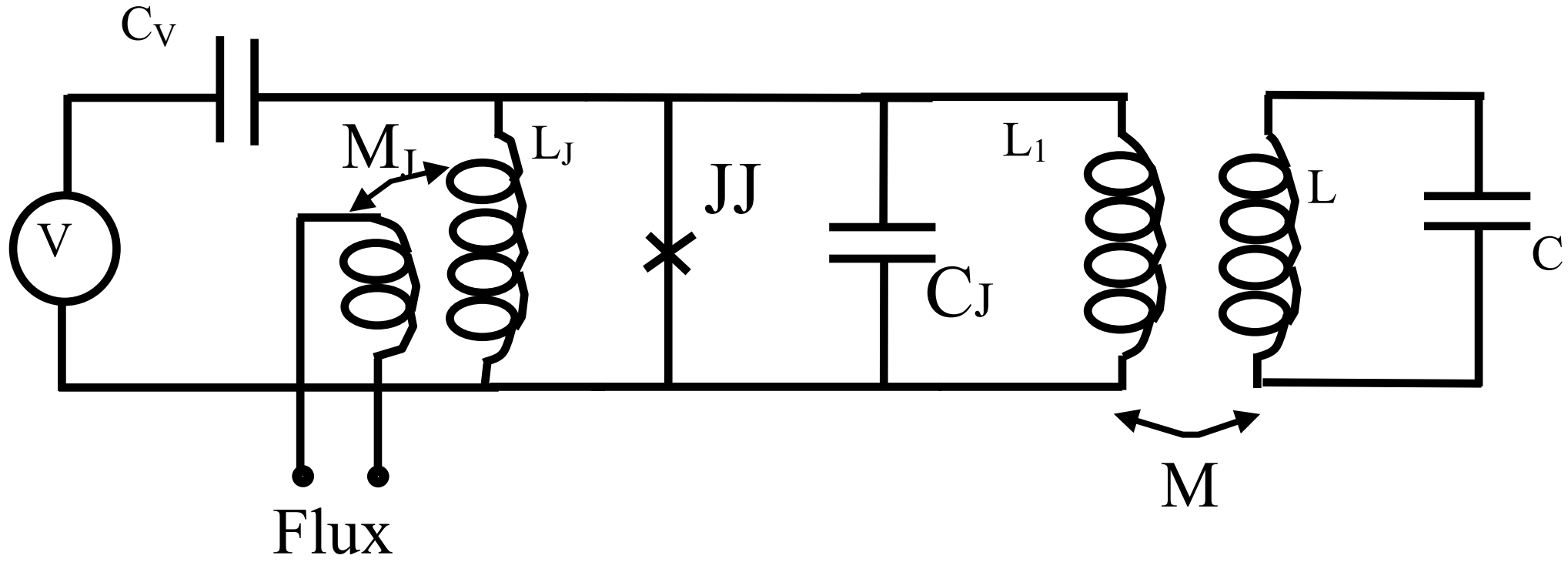

Fig. 10. The $LC$ resonator is coupled with the qubit circuit via the mutual inductance, $M$. An external *ac*-voltage, $V$, can be used for control the qubit circuit or readout of the qubit state.

A sum of the second and third terms in the right side of Eq. (1.68) can be rewritten as

$$\frac{(\varphi_J - \varphi_e')^2}{2L_J'} + const\,, \tag{1.69}$$

where,

$$\varphi_e' = \varphi_e \frac{L_1}{L_1 + L_J}, \quad L_J' = \frac{L_1 L_J}{L_1 + L_J} \tag{1.70}$$

and a "*const*" can be omitted. Then the circuit Hamiltonian can be written as a sum,

$$H = H_q + H_r + H_{dr} + H_{\text{int}}\,, \tag{1.71}$$

where,

$$H_q = \frac{q_J^2}{2(C_V + C_J)} - E_J \cos\left(\frac{2\pi}{\varphi_0}\varphi_J\right) + \frac{(\varphi_J - \varphi_e')^2}{2L_J'}\,, \tag{1.72}$$

$$H_r = \frac{\varphi^2}{2L} + \frac{q^2}{2C}, \tag{1.73}$$

$$H_{dr} = q_J \frac{VC_V}{C_V + C_J}, \tag{1.74}$$

$$H_{\text{int}} = -\frac{M}{LL_1}\varphi\varphi_J . \tag{1.75}$$

Eqs. (1.72)-(1.75) can be rewritten in terms of oscillator variables $a^+$, $a$ and Pauli spin matrices as

$$\begin{aligned}
&H_q = -\frac{\hbar\omega_q}{2}\sigma_z, \\
&H_r = \hbar\omega_r\left(a^+a + \frac{1}{2}\right) + \lambda'(a^+ + a), \\
&H_{dr} = \tilde{V}\sigma_y, \\
&H_{\text{int}} = \lambda\left(\sigma_z\cos\alpha + \sigma_x\sin\alpha\right)(a^+ + a),
\end{aligned} \tag{1.76}$$

where

$$\begin{aligned}
&\omega_r^2 = (LC)^{-1}, \\
&\tilde{V} = -\frac{\hbar V C_V}{C_V + C_J}\langle l|\frac{\partial}{\partial\varphi_J}|r\rangle, \\
&\lambda = \frac{M}{2LL_1}\sqrt{\frac{\hbar}{2\omega_r C}}(\varphi_{ll} - \varphi_{rr}), \\
&\lambda' = \lambda\frac{(\varphi_{ll} + \varphi_{rr})}{(\varphi_{ll} - \varphi_{rr})}, \\
&\varphi_{ll} = \langle l|\varphi_J|l\rangle, \\
&\varphi_{rr} = \langle r|\varphi_J|r\rangle, \\
&\cos\alpha = \frac{\varepsilon}{\sqrt{\varepsilon^2 + \Delta^2}}, \ \ \sin\alpha = \frac{|\Delta|}{\sqrt{\varepsilon^2 + \Delta^2}}.
\end{aligned} \tag{1.77}$$

We have neglected here "non-diagonal" matrix elements, $\langle l|\varphi_J|r\rangle$ and $\langle r|\varphi_J|l\rangle$, which are small because of negligible overlapping of functions, $|l\rangle$ and $|r\rangle$.

In contrast to the case of capacitive qubit-resonator interaction, we see here that the "diagonal" term (with $\sigma_z$) is present in $H_{\text{int}}$. This part of the total Hamiltonian commutes with the qubit Hamiltonian thus indicating the possibility on nondestructive readout of the qubit state. The diagonal interaction term is proportional to the parameter, $\varepsilon$, which differs from zero only in the case of an asymmetric double-well potential. There is the possibility of changing, $\varepsilon$, by variation of the external magnetic flux. The Hamiltonian (1.76) will be used below for demonstration of the approach based on adiabatic reversals for measurement of the qubit states.

In general, our approach can be described as following. Assume that we have a general Hamiltonian, which describes the qubit, the resonator, and the interaction between them:

$$\begin{aligned}
&\mathrm{H} = -A\sigma_z - B\sigma_x - C\sigma_y + \hbar\omega_r\left(a^\dagger a + \frac{1}{2}\right) + \\
&\lambda\left(A'\sigma_z + B'\sigma_x + C'\sigma_y + D'I\right)\left(a^\dagger \pm a\right) + \\
&F_q\left(t\right)\left(A''\sigma_z + B''\sigma_x + C''\sigma_y + D''I\right) + \\
&F_r\left(t\right)\left(a^\dagger \pm a\right).
\end{aligned} \tag{1.78}$$

Here the first term describes the qubit, the second term – the resonator, the third term – the qubit-resonator interaction, the fourth – the interaction between the qubit and the external field, $F_q\left(t\right)$, and the last one – interaction between the resonator and the external field $F_r\left(t\right)$. $I$ is the unitary operator.

We simplify this Hamiltonian by transferring it to the eigenfunctions of the qubit Hamiltonian to obtain, with accuracy to a constant, the qubit Hamiltonian, $-\hbar\omega_q\sigma_z/2$. Next, we transfer to the system of coordinate rotating with the qubit frequency, $\omega_q$, and ignore the fast oscillating terms in the rotating system. Finally, we assume that the external field, $F_q\left(t\right)$, acting on a qubit in the rotating frame, is either constant or oscillates with the resonator frequency, which is supposed to be small in comparison with the qubit frequency. The external field, $F_r\left(t\right)$, acting on the resonator is either absent or has the frequency equal to the resonator's frequency. Under such conditions the qubit slowly (adiabatically) oscillates between the ground and excited states. We will show that the qubit influences the resonator parameters, which allows one to determine the initial qubit state. Moreover, this measurement can be a nondestructive one. We show that the qubit may generate quasiclassical driven oscillations of the resonator or a shift the frequency of the quasiclassical resonator oscillations. In the latter case the frequency shift of the quasiclassical oscillations is proportional to the interaction parameter $\lambda$ rather than $\lambda^2$, which can be favorable for the measurement of the qubit states.

## II. Adiabatic reversals for measurement the state of a phase qubit

Below, in sections 2.1-2.7, we consider one of the methods for the detection of a state of the phase qubit. The idea is the following. The phase qubit, which can be represented by the effective spin $S = 1/2$, interacts with a resonator. (Below we omit the adjective "effective" with respect to spin.) The frequency of the resonator, $\omega_r$, is much smaller than the frequency of the qubit, $\omega_q$.The *ac* voltage, which also interacts with the phase qubit is equivalent to the application of the RF field to the spin. The frequency of the RF field itself is modulated with the frequency of the resonator. This modulation causes adiabatic reversals of the spin with the resonator frequency, $\omega_r$. These reversals drive the oscillations of the resonator. The phase of the resonator's driven oscillations depends on the spin state. Similar method has been applied in the magnetic resonance force microscopy (MRFM) for the detection of electron and nuclear spins [13,14] but has not previously been applied for the measurement of the spin state. In the MRFM experiments the amplitude of the driven vibrations rather then their phase has been detected.

In sections 2.8-2.10 we will consider the adiabatic spin reversals caused by the resonator oscillations. The back reaction of the spin on the resonator causes the resonator frequency shift, which depends on the spin state. A similar method has been used in MRFM for detection of a single electron spin [14,15]. Both techniques can be well described by the quasi-classical equations of motion if the phase qubit is in its ground or excited states.

In sections 2.11 and 2.12 we will discuss the quantum dynamics of the spin-resonator system for the case when the phase qubit is in the superposition of the ground and excited states.

### 2.1. The Hamiltonian of the spin-resonator system

In this section, we introduce an explicit dependence of the external field in (1.76),

$$\tilde{V} = \tilde{V}(t) = -\hbar\omega_R \cos(\omega_L t + \theta). \tag{2.1}$$

where $\omega_R$ is the Rabi frequency, and $\omega_L$ is the Larmor frequency of the spin, which is equivalent to the frequency of the quantum transition of the phase qubit, $\omega_q$. Then, the Hamiltonian (1.76) of the spin-resonator system can be written as

$$\begin{gathered} \mathrm{H} = -\omega_L S_z + \omega_r \left( a^\dagger a + \frac{1}{2} \right) - 2\omega_R \cos(\omega_L t + \theta) S_y + \\ \left[ 2\lambda (S_z \cos\alpha + S_x \sin\alpha) + \lambda' \right] (a^\dagger + a). \end{gathered} \tag{2.2}$$

Here and below we use the operators of the effective spin, $S_k = \sigma_k/2$, $(k = x, y, z)$, and we set $\hbar = 1$. In Eq. (2.2) the Rabi frequency describes the interaction between the qubit and the *ac* voltage, while $\lambda$ and $\lambda'$ describe the interaction between the qubit and the resonator, respectively. The phase, $\theta$, is generally time-dependent.

Our first goal is to simplify the Hamiltonian (2.2). For this we will transfer to the system of coordinates rotating with the frequency, $\omega_L$, clockwise relative to the *z*-axis. We use the rotation operator, $R_z = e^{i(\omega_L t + \theta) S_z}$. Then we transfer from operators, $S_x$ and $S_y$, to the operators

$$S_+ = S_x + iS_y,\ S_- = S_x - iS_y, \tag{2.3}$$

and use the relations,

$$R_z^\dagger S_+ R_z = e^{-i(\omega_L t + \theta)} S_+,\ R_z^\dagger S_- R_z = e^{i(\omega_L t + \theta)} S_-. \tag{2.4}$$

The Hamiltonian, $\mathrm{H}'$, in the rotating system is

$$\begin{gathered} \mathrm{H}' = R_z^\dagger \mathrm{H} R_z = -\omega_L S_z + \omega_r \left( a^\dagger a + \frac{1}{2} \right) - \omega_R \left\{ S_y \left[ 1 + \cos(2\omega_L t + 2\theta) \right] - \right. \\ \left. S_x \sin(2\omega_L t + 2\theta) \right\} + (a^\dagger + a) \left\{ \lambda' + 2\lambda S_z \cos\alpha + 2\lambda \left[ S_x \cos(\omega_L t + \theta) + S_y \sin(\omega_L t + \theta) \right] \sin\alpha \right\}. \end{gathered} \tag{2.5}$$

In the rotating system we will omit all oscillating term, assuming the averaging over fast oscillations. This procedure is justified if $\omega_R/\omega_L \ll 1$. Then, we have for the Hamiltonian $\mathrm{H}'$,

$$\mathrm{H}' = -\omega_L S_z + \omega_r\left(a^\dagger a + \frac{1}{2}\right) - \omega_R S_y + 2\lambda\left(a^\dagger + a\right) S_z \cos\alpha. \quad (2.6)$$

The wave function, $\psi'$, in the rotating system is connected to the wave function, $\psi$, in the laboratory system by the relations

$$\psi = R_z\psi', \;\; \psi' = R_z^\dagger \psi.$$

The time derivative of the operator, $R_z^\dagger$, is given by

$$\dot{R}_z^\dagger = -i\left(\omega_L + \dot{\theta}\right) S_z. \quad (2.7)$$

Using this expression we obtain the following Schrödinger equation in the rotating system:

$$i\dot{\psi}' = \left(\omega_L + \dot{\theta}\right) S_z \psi' + \mathrm{H}'\psi'. \quad (2.8)$$

Thus, we can introduce the effective Hamiltonian, $\mathrm{H}_{eff}$, which describes the spin-resonator in the rotating frame:

$$H_{eff} = \dot{\theta} S_z + \omega_r\left(a^\dagger a + \frac{1}{2}\right) - \omega_R S_y + \left(2\lambda\cos\alpha S_z + \lambda'\right)\left(a^\dagger + a\right). \quad (2.9)$$

The time derivative, $\dot{\theta}$, describes the frequency modulation of the *ac* voltage. We set

$$\dot{\theta} = -\Omega\cos\left(\omega_r t\right), \quad (2.10)$$

where $\Omega$ is the maximum deviation of the *ac* frequency from the Larmor frequency, $\omega_L$. Next, we introduce the effective interaction constant, $\lambda\cos\alpha = \Lambda$. Then, the effective Hamiltonian (2.9) takes the form,

$$H_{eff} = -\Omega\cos\left(\omega_r t\right) S_z + \omega_r\left(a^\dagger a + \frac{1}{2}\right) - \omega_R S_y + \left(2\Lambda S_z + \lambda'\right)\left(a^\dagger + a\right). \quad (2.11)$$

Note that below we will consider the adiabatic motion of the spin. This means that formally we can consider an "instantaneous" spin Hamiltonian, $-\Omega\cos\left(\omega_r t\right) S_z - \omega_R S_y$. Because of the term, $-\omega_R S_y$, the interaction Hamiltonian, $(2\Lambda S_z + \lambda')(a^\dagger + a)$, does not commute with the instantaneous spin Hamiltonian, and the nondestructive measurement of a spin state seems to be impossible. However, we are going to consider the case, $\Omega \gg \omega_r$. In this case, at times instants, $t_k = 2\pi k/\omega_r$ $(k = 0,1,2,..)$, we can formally ignore the term $-\omega_R S_y$, so that the interaction Hamiltonian will commute with the spin Hamiltonian. Below we will show that a nondestructive measurement is possible if the measurement process starts and ends at times $t_k = 2\pi k/\omega_r$.

### 2.3. Heisenberg equations of motion

In this section we obtain the Heisenberg equations of motion for the qubit-resonator system. We will use the operator of the dimensionless effective flux,

$$\varphi_r = \frac{1}{\sqrt{2}}\left(a^\dagger + a\right), \tag{2.12}$$

and the operator of dimensionless charge,

$$q_r = \frac{i}{\sqrt{2}}\left(a^\dagger - a\right). \tag{2.13}$$

as the time-dependent operators describing the resonator. The evolution of a qubit will be described by the time-dependent spin operators. The connections between the dimensionless operators here and the dimensional operators of the resonator in (1.73) are

$$\varphi = -\left(\frac{\hbar}{C\omega_r}\right)^{1/2} \varphi_r,\ q = -\left(\hbar C\omega_r\right)^{1/2} q_r. \tag{2.14}$$

Note that $\varphi$ is an effective flux which is connected to the voltage, $V$, in the resonator as

$$\varphi(t) = \int_{-\infty}^{t} V(t')dt'. \tag{2.15}$$

In terms of new operators, the effective Hamiltonian (2.11) becomes

$$\mathrm{H}_{eff} = -\Omega\cos(\omega_r t) S_z + \frac{\omega_r}{2}\left(\varphi_r^2 + q_r^2\right) - \omega_R S_y + \sqrt{2}\left(2\Lambda S_z + \lambda'\right)\varphi_r. \tag{2.16}$$

We use the standard relations for commutators,

$$\left[\varphi_r, q_r\right] = i,\ \left[S_j, S_k\right] = i\varepsilon_{jkm} S_m, \tag{2.17}$$

where $\varepsilon_{jkm}$ is the antisymmetric tensor with $\varepsilon_{xyz=1}$.
The equation of motion for the operator $\vec{S}=\left(S_x,S_y,S_z\right)$ is

$$\dot{\vec{S}}=i\left[\mathrm{H}_{eff},\vec{S}\right]=\left[\vec{S}\times\vec{B}\right], \tag{2.18}$$

where $\vec{B}$ is the vector operator describing the effective field acting on the spin. The components of the operator $\vec{B}$ are

$$B_x=0,\ B_y=\omega_R,\ B_z=\Omega\cos\left(\omega_r t\right)-2\sqrt{2}\Lambda\varphi_r. \tag{2.19}$$

In a similar way we obtain the equations of motion for the operators, $\varphi_r$ and $q_r$ :

$$\dot{\varphi}_r=\omega_r q_r,\ \dot{q}_r=-\omega_r\varphi_r-\sqrt{2}\left(2\Lambda S_z+\lambda'\right). \tag{2.20}$$

These two operator equations can be reduced to a single second order operator equation for the flux, $\varphi_r$,

$$\ddot{\varphi}_r+\omega_r^2\varphi_r=f, \tag{2.21}$$

where $f$ is the operator describing the effective "external force" provided by spin which drives the resonator:

$$f=-\sqrt{2}\omega_r\left(2\Lambda S_z+\lambda'\right). \tag{2.22}$$

The constant force, $-\sqrt{2}\omega_r\lambda'$, changes the equilibrium value of the flux, $\varphi_r$, from zero to $\varphi_{r,eq}=-\sqrt{2}\,\lambda'/\omega_r$. The new equilibrium flux is the flux in the resonator produced by the electric current in the qubit system. If we count a flux from the new equilibrium value,

$$\varphi_r\to\varphi_r+\varphi_{r,eq},$$

we will get the equation of motion (2.21) with the force

$$f=-2\sqrt{2}\Lambda\omega_r S_z.$$

## 2.4. Classical equations of motion

Assume that the adiabatic regime is realized. Namely, the resonator's frequency is much smaller than the Rabi frequency, which itself is small compared to the Larmor frequency,

$$\omega_r \ll \omega_R \ll \omega_L \,. \tag{2.23}$$

In this case, the *z*-component of the effective field in (2.19) will change so slowly that the angle between the average spin, $\langle \vec{S} \rangle$, and the average field, $\langle \vec{B} \rangle$, will not change (i.e., is an adiabatic invariant).
Assume that the average spin points in the direction of the vector $\langle \vec{B} \rangle$, or in the opposite direction. In both cases the spin drives the resonator's oscillations with a definite phase. If so, there is no way for the formation of the resonator's Schrödinger cat-state, and our system can be described by the quasiclassical equations of motion.
In this case, we take the quantum-mechanical averages of the Heisenberg equations (2.18) and (2.21), and ignore the quantum correlations, $C_{SB}(t)$, between the resonator and qubit operators in Eq. (2.18):

$$C_{SB}^{(k,j)}(t) = \langle S_k B_j \rangle - \langle S_k \rangle \langle B_j \rangle \to 0. \tag{2.24}$$

Note, that the validity of the assumption (2.24) can be verified by numerical simulations. Below we omit the angular brackets when writing the equations for averages,

$$\ddot{\varphi}_r + \omega_r^2 \varphi_r = f, \; \dot{\vec{S}} = \left[ \vec{S} \times \vec{B} \right], \tag{2.25}$$

where

$$B_x = 0, \; B_y = \omega_R, \; B_z = \Omega \cos(\omega_r t) - 2\sqrt{2}\Lambda \varphi_r, \; f = -2\sqrt{2}\Lambda \omega_r S_z. \tag{2.26}$$

These equations for averages look like the corresponding equations for the operators.

### 2.5. Adiabatic reversals of the qubit

Because of the inequalities (2.23), the spin (qubit), which initially points in (or opposite to) the direction of the effective field (2.26), will follow the effective field. The approximate (adiabatic) solution of the spin equation in (2.25) can be found if we set $\dot{\vec{S}} = 0$. In this case, we have

$$S_z = \pm S B_z / B, \; S_y = \pm S B_y / B, \; S_x = 0, \tag{2.27}$$

where $B$ is the magnitude of the effective field,

$$B = \left( B_y^2 + B_z^2 \right)^{1/2}. \tag{2.28}$$

S in (2.27) is the magnitude of spin $(S = 1/2)$, and the signs $\pm$ correspond to the initial qubit state, $|0\rangle$ or $|1\rangle$ $(S_z = \pm 1/2)$.

Below in this section, we assume that the action of the resonator on the qubit can be ignored compared to the action of the *ac* voltage. This implies that

$$\Omega \gg 2\sqrt{2}\Lambda|\varphi_r|. \tag{2.29}$$

In this case, the *z*-component of the effective field is given by

$$B_z \approx \Omega\cos(\omega_r t). \tag{2.30}$$

In this approximation, the component, $B_z$, changes between $\Omega$ and $-\Omega$, while the y-component, $B_y = \omega_R$, remains constant. The change of the direction of the effective field is shown schematically in Fig. 11. Note that we assume

$$\Omega \gg \omega_R. \tag{2.31}$$

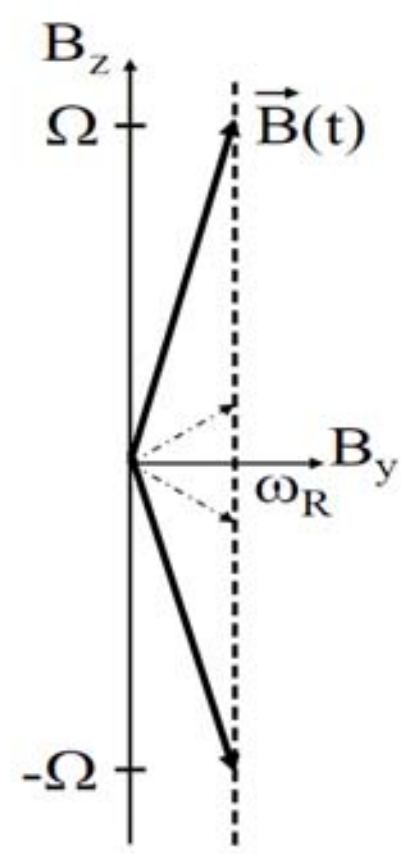

Fig. 11: Change in the direction of the effective field, $\vec{B}$. Arrow shows the vector $\vec{B}$. Dashed line indicates the constant value, $B_y = \omega_R$.

In this case, the frequency modulation of the *ac* voltage provides, approximately, the periodic reversals of the spin.

Taking into account our approximation (2.30), the *z*-component of the spin becomes

$$S_z = \pm\frac{\Omega}{2}\frac{\cos(\omega_r t)}{\sqrt{\omega_R^2 + (\Omega\cos(\omega_r t))^2}}. \tag{2.32}$$

## 2.6. Oscillations of the resonator

Now we consider the driven oscillations of the resonator. From Eqs. (2.25), (2.26) we have

$$\ddot{\varphi}_r + \omega_r^2 \varphi_r = f,\ f = -2\sqrt{2}\Lambda\omega_r S_z. \qquad (2.33)$$

The expression for the effective force can be written as

$$f = f_0 \cos(\omega_r t), \qquad (2.34)$$

where the "amplitude of the force", $f_0$, is given by

$$f_0 = \mp \frac{\sqrt{2}\Lambda\omega_r\Omega}{\sqrt{\omega_R^2 + \left(\Omega\cos(\omega_r t)\right)^2}}. \qquad (2.35)$$

The upper sign "-" corresponds to the initial ground state of qubit, $|0\rangle$, and the lower sign "+" corresponds to the excited state, $|1\rangle$.

The solution of Eq. (2.33) for $\varphi_r$ describes a generation of the driven oscillations in the resonator. In order to obtain the approximate solution of this equation we will express $\cos^2(\omega_r t)$ in the denominator of the force amplitude (2.35) as

$$\cos^2(\omega_r t) = \frac{1}{2}\left[1 + \cos(2\omega_r t)\right]. \qquad (2.36)$$

If we ignore the fast oscillating term, $\cos(2\omega_r t)$, the expression for the force amplitude becomes time-independent:

$$f_0 = \mp \frac{2\Lambda\omega_r\Omega}{\sqrt{\Omega^2 + 2\omega_R^2}}. \qquad (2.37)$$

Taking into consideration that $\Omega \gg \omega_R$, we can reduce this expression to

$$f_0 = \mp 2\Lambda\omega_r. \qquad (2.38)$$

Now, we can obtain the solution of Eq. (2.33), which describes the driven oscillations of the resonator with the increasing amplitude:

$$\varphi_r(t) = \mp \Lambda t \sin(\omega_r t). \qquad (2.39)$$

Consider the phase shift of the flux oscillations, $\varphi_r(t)$, with respect to the phase of the frequency oscillations of the voltage. The oscillations of the frequency of the *ac* voltage are given by the expression (2.10): $\dot{\theta} = -\Omega\cos(\omega_r t)$. We can rewrite the driven oscillations of the flux in (2.39) in the form

$$\varphi_r(t) = -\Lambda t \cos\left(\omega_r t \pm \frac{\pi}{2}\right). \tag{2.40}$$

Thus, for the ground state, $|0\rangle$, of the qubit, the phase of the driven oscillations is shifted by $-\pi/2$, while for the excited state, $|1\rangle$, is shifted by $\pi/2$. By measuring the phase of the driven oscillations of the resonator one can determine the initial state of the qubit. Note, that the described measurement can be implemented as a nondestructive measurement: after an integer number of the periods of the resonator, $T_r = 2\pi/\omega_r$, the qubit returns to its initial state.

## 2.7. Conditions for the adiabatic reversals

In this section, we will obtain the conditions for the adiabatic reversals of the spin. The spin follows the effective field if the angular speed of the effective field rotations is much smaller than the frequency of spin precession about this field. The angular speed of the effective field is the rate of change of its polar angle, which we denote as $\theta_p$. Taking into consideration the inequality (2.29), we have the following expressions for the effective field:

$$B_x = 0,\ B_y = \omega_R,\ B_z = \Omega\cos(\omega_r t). \tag{2.41}$$

Thus, the polar angle of the effective field is defined by the expressions

$$\cos\theta_p = \frac{\Omega\cos(\omega_r t)}{\sqrt{\omega_R^2 + (\Omega\cos(\omega_r t))^2}},\ \sin\theta_p = \frac{\omega_R}{\sqrt{\omega_R^2 + (\Omega\cos(\omega_r t))^2}}. \tag{2.42}$$

Taking a time derivative from the second equation in (2.42), and using the first equation, we obtain

$$\dot{\theta}_p = \frac{\omega_R \omega_r \Omega \sin(\omega_r t)}{\omega_R^2 + \Omega^2\cos^2(\omega_r t)}. \tag{2.43}$$

One can see that the maximum angular speed,

$$\dot{\theta}_p = \frac{\omega_r \Omega}{\omega_R}, \tag{2.44}$$

takes place at $\omega_r t = \pi/2$, when the spin passes the *x*-*y*-plane.

The frequency of the spin precession about the effective field is

$$\left(\omega_R^2 + \Omega^2\cos^2(\omega_r t)\right)^{1/2}.$$

This frequency takes its minimum value at the same time, $\omega_r t = \pi/2$, when the angular speed, $\dot{\theta}_p$, takes its maximum value. Thus, the condition for the adiabatic spin reversals can be written as

$$\frac{\omega_r \Omega}{\omega_R} \ll \omega_R, \tag{2.45}$$

or $\omega_r \Omega \ll \omega_R^2$. As we assumed in (2.31), that $\Omega \gg \omega_R$, it follows that the condition (2.23), $\omega_r \ll \omega_R$, is not sufficient. We must require a stronger inequality for the resonator frequency,

$$\omega_r \ll \omega_R^2/\Omega, \tag{2.46}$$

in order to satisfy the condition for adiabatic reversals of the spin.

### 2.8. Adiabatic reversals of the spin caused by the resonator oscillations

In this section we consider the regime when the frequency modulation of the *ac* voltage is not employed: $\Omega = 0$ in Eq. (2.26). Instead, one set the oscillations of the resonator. As an example, we assume for definiteness that $\Lambda > 0$, and set the initial conditions in Eq. (2.25):

$$\varphi_r(0) = -A, \ \dot{\varphi}_r(0) = 0. \tag{2.47}$$

In this case, the adiabatic reversals of the spin will be caused by the oscillations in the resonator if the following condition is satisfied:

$$2\sqrt{2}\Lambda A \gg \omega_R. \tag{2.48}$$

The *z*-component of the effective field will oscillate between the values $\pm 2\sqrt{2}\Lambda A$, while the *y*-component again remains constant (see Fig. 12).

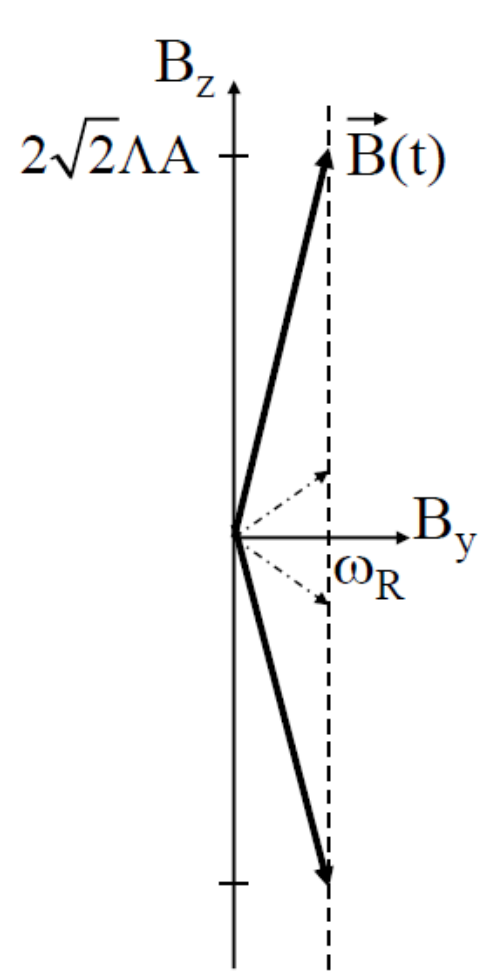

Fig. 12: Adiabatic reversals of the effective field caused by the oscillations in the resonator.

We expect that the back action of the spin on the resonator causes the resonator frequency shift, which depends on the initial state of the qubit.

The adiabatic solution of Eq. (2.25) for a spin is given by Eq. (2.27). The *z*-component of the spin can be written as

$$S_z = \mp \frac{\sqrt{2}\Lambda\varphi_r}{\sqrt{\omega_R^2 + 8\Lambda^2\varphi_r^2}}. \tag{2.49}$$

The flux oscillations in the resonator are described by the first Eq. (2.25),

$$\ddot{\varphi}_r + \omega_r^2\varphi_r = f, \tag{2.50}$$

where the effective force is given by

$$f = -2\sqrt{2}\Lambda\omega_r S_z = \pm \frac{4\Lambda^2\omega_r\varphi_r}{\sqrt{\omega_R^2 + 8\Lambda^2\varphi_r^2}}. \tag{2.51}$$

Eq. (2.50) with the force (2.51) represents a nonlinear equation for the resonator oscillations, $\varphi_r$. We are looking for the approximate solution of Eq. (2.50) in the form

$$\varphi_r = -A\cos(\omega t), \tag{2.52}$$

where the resonator frequency, $\omega$, is slightly shifted from its unperturbed value, $\omega_r$ due to the resonator-spin interaction. Substituting (2.52) into the equation for $\varphi_r$ (2.50), we obtain

$$A\left(\omega^2 - \omega_r^2\right)\cos(\omega t) = \mp \frac{4\Lambda^2\omega_r A\cos(\omega t)}{\sqrt{\omega_R^2 + 8\Lambda^2 A^2\cos^2(\omega t)}}. \tag{2.53}$$

Again, we use the expression (2.36) and ignore the fast oscillating term, $(1/2)\cos(2\omega_r t)$. Then we obtain the equation for the resonator frequency, $\omega$,

$$\omega^2 = \omega_r^2 \mp \frac{4\Lambda^2\omega_r}{\sqrt{\omega_R^2 + 4\Lambda^2 A^2}}. \tag{2.54}$$

Note, that for small values of the resonator's oscillating amplitude $\left(4\Lambda^2 A^2 \ll \omega_R^2\right)$ the resonator frequency shift is proportional to the square of the interaction constant, $\Lambda^2$. However, for large values of $A$, the frequency shift is proportional to $\Lambda$. The quasi-classical resonator oscillations with relatively large amplitudes can be easily detected, which may become an important advantage in the experimental implementation of a single qubit measurement.

Taking into consideration the inequality (2.48), we reduce this expression to

$$\omega^2 = \omega_r^2 \mp 2\Lambda\omega_r / A. \tag{2.55}$$

Assuming that the frequency shift is small, we have from (2.55)

$$\omega = \omega_r \mp \Lambda / A. \tag{2.56}$$

Thus, the frequency shift of the resonator is given by

$$\delta\omega = \mp\Lambda / A. \tag{2.57}$$

The frequency shift is negative for the ground state of the qubit, $|0\rangle$, and positive for the excited state, $|1\rangle$. Thus, by measuring the resonator frequency shift one can measure the initial state of the qubit. Note, that this measurement can also be implemented as a nondestructive measurement: after an integer number of the resonator oscillations, the qubit returns to its initial state.

### 2.9. More accurate estimate for the frequency shift

In this section we will derive a more accurate solution of Eq. (2.50) using a regular perturbation theory developed by Bogolubov and Mitropolsky [16]. For this, we will rewrite Eqs. (2.50), (2.51) in the standard form,

$$\ddot{\varphi}_r + \varphi_r = \varepsilon \tilde{f}(\varphi_r). \tag{2.58}$$

Here we introduced a dimensionless time,

$$\tau = \omega_r t, \tag{2.59}$$

and take a derivative with respect to $\tau$ : $\ddot{\varphi}_r = d^2\varphi_r / d\tau^2$ . We define $\tilde{f}(\varphi_r)$ as

$$\tilde{f}(\varphi_r) = \frac{2\sqrt{2}\Lambda\varphi_r}{\sqrt{\omega_R^2 + 8\Lambda^2\varphi_r^2}}, \tag{2.60}$$

and the value of a small parameter, $\varepsilon$ , is

$$\varepsilon = \pm\frac{\sqrt{2}\Lambda}{\omega_r}. \tag{2.61}$$

We look for a solution of Eq. (2.58) to the first order on $\varepsilon$, in the form

$$\varphi_r(\tau) = a(\tau)\cos\psi(\tau). \tag{2.62}$$

The functions, $a(\tau)$ and $\psi(\tau)$, satisfy the equations

$$\begin{aligned}
&\dot{a} = \varepsilon P(a),\\
&\dot{\psi} = 1 + \varepsilon Q(a),\\
&P(a) = -\frac{1}{2\pi}\int_0^{2\pi} \tilde{f}(a\cos\psi)\sin\psi d\psi,\\
&Q(a) = -\frac{1}{2\pi a}\int_0^{2\pi} \tilde{f}(a\cos\psi)\cos\psi d\psi.
\end{aligned} \tag{2.63}$$

Substituting the expression for $\tilde{f}(a\cos\psi)$,

$$\tilde{f}(a\cos\psi) = \frac{2\sqrt{2}\Lambda a\cos\psi}{\sqrt{\omega_R^2 + 8\Lambda^2 a^2\cos^2\psi}}, \tag{2.64}$$

into (2.63), we obtain

$$\begin{aligned}
&P(a) = 0,\\
&Q(a) = -\frac{4\sqrt{2}\Lambda}{\pi k^2\sqrt{\omega_R^2 + 8\Lambda^2 a^2}}\left[\left(k^2 - 1\right)K(k) + E(k)\right].
\end{aligned} \tag{2.65}$$

In this expression,

$$k^2 = \frac{8\Lambda^2 a^2}{\omega_R^2 + 8\Lambda^2 a^2}, \tag{2.66}$$

and $K(k)$ and $E(k)$ are the complete elliptical integrals of the first and second kind, respectively. Taking into consideration the inequality (2.48), we obtain $k \approx 1$. Then, we have

$$\left(k^2 - 1\right)K(k) + E(k) \approx 1, \tag{2.67}$$

and, consequently, $Q(a) \approx -2/\pi A$.

The solution of Eqs. (2.63) can be written as

$$a = A,$$
$$\dot{\psi} = 1 \mp \frac{2\sqrt{2}\Lambda}{\pi\omega_r A}. \tag{2.68}$$

Returning to the dimensional time, we have

$$\frac{d\psi}{dt} = \omega_r \mp \delta\omega,\ \delta\omega = \frac{2\sqrt{2}\Lambda}{\pi A}. \tag{2.69}$$

The frequency of the resonator is $\omega = d\psi/dt$. Thus, this more accurate computation reveals an additional factor,

$$\frac{2\sqrt{2}}{\pi} \approx 0.9, \tag{2.70}$$

in the expression for the frequency shift, $\delta\omega$, (compare Eq. (2.57) with Eq. (2.69)).

### 2.10. Conditions for the adiabatic reversals

In this section we present the conditions for the adiabatic spin reversals caused by the resonator oscillations. The effective field (2.26) on the spin with no frequency modulation $(\Omega = 0)$ is given by

$$B_x = 0,\ B_y = \omega_R,\ B_z = -2\sqrt{2}\Lambda\varphi_r. \tag{2.71}$$

The polar angle of the effective field in this case is given by the expression

$$\cos\theta_p = -\frac{2\sqrt{2}\Lambda\varphi_r}{\sqrt{\omega_R^2 + 8\Lambda^2\varphi_r^2}},\ \sin\theta_p = \frac{\omega_R}{\sqrt{\omega_R^2 + 8\Lambda^2\varphi_r^2}}. \tag{2.72}$$

Taking the time derivative from the second equation in (2.72) and substituting $\cos\theta_p$ from the first equation, we have

$$\dot{\theta}_p = \frac{2\sqrt{2}\Lambda\omega_R\dot{\varphi}_r}{\omega_R^2 + 8\Lambda^2\varphi_r^2}. \tag{2.73}$$

We use Eq. (2.52) for the flux, $\varphi_r$, to obtain

$$\dot{\theta}_p = \frac{2\sqrt{2}\Lambda\omega_R\omega A\sin(\omega t)}{\omega_R^2 + 8\Lambda^2 A^2\cos^2(\omega t)}. \tag{2.74}$$

The maximum value of $\dot{\theta}_p$ is

$$\dot{\theta}_p = 2\sqrt{2}\Lambda\omega A/\omega_R, \tag{2.75}$$

at $\omega t = \pi/2$, i.e. when the effective field passes through the transversal $x$-$y$-plane.

The frequency of the spin precession about the effective field,

$$\sqrt{\omega_R^2 + 8\Lambda^2 A^2 \cos^2(\omega t)}, \tag{2.76}$$

has the minimum value, $\omega_R$, at the same transversal plane $(\omega t = \pi/2)$.

For adiabatic reversals to be implemented, the value of $\dot{\theta}_p$ must always be much smaller than the frequency of spin precession about the effective field. Thus, the condition for adiabatic reversals is

$$2\sqrt{2}\Lambda\omega A/\omega_R \ll \omega_R. \tag{2.77}$$

Note that we assume a small frequency shift of the resonator, so that $\omega \approx \omega_r$. From Eq. (2.48) we have $2\sqrt{2}\Lambda A \gg \omega_R$. This means that the condition (2.23), $\omega_r \ll \omega_R$, is not sufficient. It follows from Eq. (2.77) that the resonator frequency must satisfy a stronger inequality:

$$\omega_r \ll \omega_R^2 / 2\sqrt{2}\,\Lambda A. \tag{2.78}$$

**2.11. Quantum effects in the resonator's dynamics**

So far we have used quasiclassical equations of motion (2.25), (2.26). These equations are valid when the spin points in (or opposite to) the direction of the effective field. In our analysis the initial effective field on the spin points in the direction which is very close to the positive $z$-direction (i.e., the polar angle, $\theta_p(0)$, is close to zero). This means that the initial state of the qubit is $|0\rangle$ or $|1\rangle$.

In general, the initial state of the qubit is a superpositional state,

$$c_0|0\rangle + c_1|1\rangle, \tag{2.79}$$

where, $c_0$ and $c_1$, are the complex amplitudes of the qubit states. The superpositional qubit state generates the superposition of the effective forces on the resonator - a situation which cannot be described by the quasiclassical equations of motion.

Then, we have to transfer to the Schrödinger equation,

$$i\dot{\Psi} = \mathrm{H}\Psi, \tag{2.80}$$

where $\Psi$ is a spinor,

$$\Psi = \Psi(\varphi_r, t) = \begin{pmatrix} \psi_1(\varphi_r, t) \\ \psi_2(\varphi_r, t) \end{pmatrix}, \tag{2.81}$$

and $\varphi_r$ is the effective dimensionless flux which is now an input of the spinor components, $\psi_1$ and $\psi_2$.

In order to describe the initial quasiclassical state of the resonator, we use the coherent state, $u_\alpha(\varphi_r)$, which is the eigenfunction of the operator, $a$,

$$u_\alpha(\varphi_r) = \pi^{-1/4} \exp\left[-\left(\varphi_r/\sqrt{2} - \alpha\right)^2 + \left(\alpha^2 - |\alpha|^2\right)/2\right]. \tag{2.82}$$

The average value of the dimensionless flux and the charge can be expressed in terms of the parameter, $\alpha$,

$$\langle \varphi_r \rangle = \frac{\alpha^* + \alpha}{\sqrt{2}}, \ \langle q_r \rangle = i\frac{\alpha^* - \alpha}{\sqrt{2}}. \tag{2.83}$$

We will represent the initial spinor, $\Psi(\varphi_r, 0)$, as a product of the flux and spin components:

$$\Psi(\varphi_r, 0) = u_\alpha(\varphi_r)\chi(0), \ \chi(0) = \begin{pmatrix} c_0 \\ c_1 \end{pmatrix}. \tag{2.84}$$

The system we are considering is equivalent to the spin-cantilever system in the MRFM context [14]. Thus, we will use the results of the numerical simulations from [14] in order to describe the quantum behavior of the resonator interacting with the phase qubit.

According to the results of the numerical simulations, the spinor, $\Psi(\varphi_r, t)$, can be approximately represented as a sum of two spinors,

$$\Psi(\varphi_r, t) \approx c_0 \psi_a(\varphi_r, t)\chi_a(t) + c_1 \psi_b(\varphi_r, t)\chi_b(t), \tag{2.85}$$

where $\chi_a(t)$ describes the spin which points in the direction of the effective field, while $\chi_b(t)$ describes the spin which points in the direction opposite to the effective field.

In the case of the adiabatic reversals caused by the frequency modulation of the *ac* voltage, the spinors, $\chi_a(t)$ and $\chi_b(t)$, are approximately the two eigenfunctions of the operator $\langle \vec{B}_{eff} \rangle \cdot \vec{S}$, where $\langle \vec{B}_{eff} \rangle$ is given by Eq. (2.41). The normalized functions, $\psi_a(\varphi_r, t)$ and $\psi_b(\varphi_r, t)$, describe the two peaks, which oscillate with an increasing amplitude. The positions of the peaks can be described approximately by Eq. (2.40), where the upper sign refers to $\psi_a(\varphi_r,t)$, and the lower sign refers to $\psi_b(\varphi_r,t)$. (See qualitative picture in Fig. 13.)

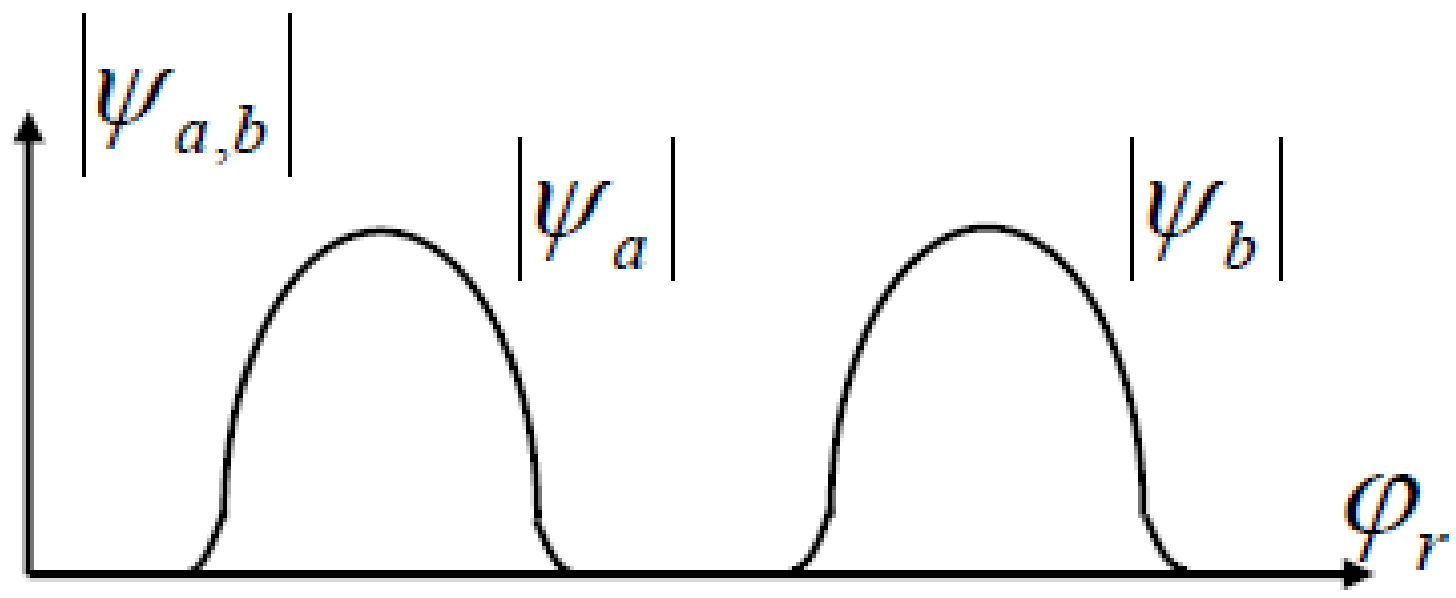

Fig. 13: Qualitative form of the functions, $|\psi_a(\varphi_r,t)|$ and $|\psi_b(\varphi_r,t)|$, at a fixed time, *t*.

Fig. 13 describes the Schrödinger cat state for the resonator: two macroscopically distinguishable values of flux at the same time. In a real experiment, the Schrödinger cat state quickly collapses, and one obtains the state, $\psi_a(\varphi_r,t)\chi_a(t)$, with the probability, $|c_0|^2$, or the state, $\psi_b(\varphi_r,t)\chi_b(t)$, with the probability, $|c_1|^2$.

A similar situation takes place when the spin adiabatic reversals are caused by the resonator's oscillations. In this case, the spinor, $\Psi(\varphi_r,t)$, can be again represented as a superposition, (2.85). The spatial functions, $\psi_a(\varphi_r,t)$ and $\psi_b(\varphi_r,t)$, now describe two peaks oscillating with approximately constant amplitudes. The frequencies of the peaks are given, approximately, by expressions (2.69). The spinor, $\chi_a(t)$, is approximately the eigenfunction of the operator, $\vec{B}_a \cdot \vec{S}$, with the eigenvalue $+1/2$. The effective field, $\vec{B}_a = \vec{B}_a(t)$ is given by the expression (2.71), where $\varphi_r = \varphi_{r,a}(t)$ is the position of the peak described by the spatial function, $\psi_a(\varphi_r,t)$. Correspondingly, the spinor, $\chi_b(t)$, is approximately the eigenfunction of the operator, $\vec{B}_b \cdot \vec{S}$, with the eigenvalue $-1/2$. The effective field, $\vec{B}_b = \vec{B}_b(t)$ is given by the expression (2.71), where $\varphi_r = \varphi_{r,b}(t)$ is the position of the peak described by the spatial function, $\psi_b(\varphi_r,t)$. Note, that unlike the case of the frequency modulation, we have now two different effective fields, $\vec{B}_a(t)$ and $\vec{B}_b(t)$. Correspondingly, the spin functions, $\chi_a(t)$ and $\chi_b(t)$, are not orthogonal to each other, except for the instants when the two peaks in Fig. 13 overlap. However, the physical picture in this case is similar to the case of the frequency modulation.

Again we have the Schrödinger cat state (see Fig. 13), which describes the two macroscopically distinguishable fluxes existing at the same time. One value of the flux corresponds to the spin pointing in the direction of the effective spin, and the other one corresponds to the spin direction opposite to the effective field. The collapse of the Schrödinger cat state preserves one of the two peaks and the corresponding spin state, and eliminates the other one.

## 2.12. Interaction between the resonator and the environment

The interaction between the resonator and the thermal environment in our problem is also equivalent to the corresponding interaction between the cantilever and the environment in the MRFM context [14]. Again, this allows us to use the results of the computer simulations for the MRFM [14], and to describe qualitatively the effects of the environment.

The mixed state of the resonator-qubit system can be described by the reduced density matrix

$$\rho=\rho\left(\varphi_r,\varphi_r',t\right)=\begin{pmatrix}\rho_{\frac{1}{2},\frac{1}{2}}\left(\varphi_r,\varphi_r',t\right) & \rho_{\frac{1}{2},-\frac{1}{2}}\left(\varphi_r,\varphi_r',t\right)\\ \rho_{-\frac{1}{2},\frac{1}{2}}\left(\varphi_r,\varphi_r',t\right) & \rho_{-\frac{1}{2},-\frac{1}{2}}\left(\varphi_r,\varphi_r',t\right)\end{pmatrix}. \tag{2.86}$$

A function, $\rho_{s,s'}\left(\varphi_r,\varphi_r',t\right)$, where $s,s'=\pm 1/2$, describes the following: If $s=s'$ and $\varphi_r=\varphi_r'$, the corresponding function gives the probability that the effective spin has the value, $s$, and the probability density that the flux has the value, $\varphi_r$. In general, the function, $\rho_{s,s'}\left(\varphi_r,\varphi_r',t\right)$, describes the situation when the effective spin has at the same time the values, $s$ and $s'$, and the flux has the values, $\varphi_r$ and $\varphi_r'$ (Schrödinger cat state). The initial reduced density matrix can be taken as the product of the flux and the spin parts:

$$\rho\left(\varphi_r,\varphi_r',0\right)=u_\alpha\left(\varphi_r\right)u_\alpha^\dagger\left(\varphi_r'\right)\begin{pmatrix}\left|c_0\right|^2 & c_0c_1^*\\ c_0^* & \left|c_1\right|^2\end{pmatrix}. \tag{2.87}$$

The equation of motion for the density matrix can be written as

$$\frac{\partial\rho}{\partial\tau}=\left[\mathrm{H}_{eff},\rho\right]-\frac{1}{Q}\left[\frac{1}{2}\left(\varphi_r-\varphi_r'\right)\left(\partial_{\varphi_r}-\partial_{\varphi_r'}\right)+D\left(\varphi_r-\varphi_r'\right)^2\right]\rho, \tag{2.88}$$

where $\mathrm{H}_{eff}$ is the effective Hamiltonian in (2.16), $Q$ is the quality factor of the resonator, $D=k_BT/\hbar\omega_r$, and $\partial_{\varphi_r}$ means $\partial/\partial\varphi_r$. Note that this equation takes into consideration the interaction between the resonator and the environment, and ignores the direct interaction between the qubit and the environment.

The numerical simulations show that the reduced density matrix, $\rho_{s,s'}\left(\varphi_r,\varphi_r',t\right)$, can be approximately represented as a sum of four terms:

$$\rho=\rho^{(1)}+\rho^{(2)}+\rho^{(3)}+\rho^{(4)}. \tag{2.89}$$

The matrices, $\rho^{(1)}$ and $\rho^{(2)}$, describe the peaks oscillating along the diagonal on the $\varphi-\varphi'$– plane. The matrices, $\rho^{(3)}$ and $\rho^{(4)}$, describe the non-diagonal peaks (see Fig. 14).

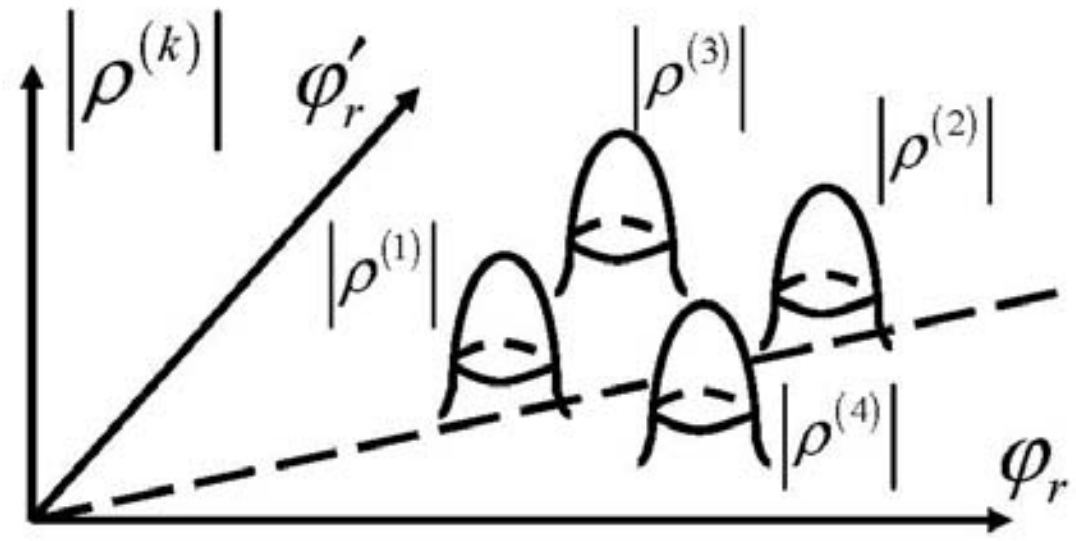

Fig. 14: Four peaks corresponding to the matrices, $\rho^{(k)}$, in (2.89).

The diagonal matrices, $\rho^{(1)}$ and $\rho^{(2)}$, can be approximately represented as the product of the flux and spin parts:

$$\rho^{(1)} = \hat{\psi}_a\left(\varphi_r, \varphi_r', t\right)\hat{\chi}_a\left(t\right),\ \rho^{(2)} = \hat{\psi}_b\left(\varphi_r, \varphi_r', t\right)\hat{\chi}_b\left(t\right). \quad (2.90)$$

The spin density matrices, $\hat{\chi}_a\left(t\right)$ and $\hat{\chi}_b\left(t\right)$, describe the "pure" spin states. They can be constructed from the corresponding spinors in (2.85):

$$\hat{\chi}_a = \chi_a\chi_a^\dagger,\ \hat{\chi}_b = \chi_b\chi_b^\dagger. \quad (2.91)$$

These matrices describe the spin which points in (or opposite to) the direction of the effective field, exactly as was described above for the Schrödinger dynamics. The functions, $\hat{\psi}_a\left(\varphi_r, \varphi_r', t\right)$ and $\hat{\psi}_b\left(\varphi_r, \varphi_r', t\right)$, describe the oscillations of the flux in the resonator corresponding to the two directions of the spin. For large values of the quality factor, the flux oscillations are exactly the same as they were in the Schrödinger dynamics.

The two non-diagonal peaks which describe the coherence between the two macroscopically distinct values of the flux (the Schrödinger cat state) quickly diminish due to the interaction with the thermal environment. The disappearance of the non-diagonal peaks reflects the collapse of the superpositional state, $\Psi\left(\varphi_r, t\right)$, into the states $\psi_a\left(\varphi_r, t\right)\chi_a\left(t\right)$ or $\psi_b\left(\varphi_r, t\right)\chi_b\left(t\right)$. Each of these two states describes a definite direction of the effective spin with respect to the effective field (parallel or antiparallel). Each of these states also describes a definite value of the flux, $\varphi_r\left(t\right)$, at time $t$ (with accuracy to the quantum uncertainty of the flux).

## Conclusion

Our work here describes the basic properties of the phase qubit and its interaction with a microstrip resonator. We considered the Hamiltonians of the phase qubit for the cases of capacitive and inductive couplings. It was shown that in second order perturbation theory the qubit-resonator interaction allows one to perform a nondestructive measurement of the phase qubit. In the highest order perturbation theory, the resonator spectrum becomes non-equidistant. In the case of the inductive coupling, the qubit-resonator interaction is similar to that for magnetic resonance force microscopy (MRFM). We have derived the quasi-classical equations of motion for the qubit-resonator system and proposed using an adiabatic approach, similar to one used in the MRFM, for the measurement of the phase qubit state. We have considered two methods: one based on the measurement of the phase of the driven resonator's oscillations; and the other one based on the measurement of the resonator frequency shift. Both methods allow a nondestructive measurement of the qubit state. The conditions for adiabatic dynamics have been formulated. Quantum effects and interactions with the environment have also been discussed.

**Acknowledgement**

This work was carried out under the auspices of the National Nuclear Security Administration of the U.S. Department of Energy at Los Alamos National Laboratory under Contract No. DE-AC52-06NA25396 and by Lawrence Livermore National Laboratory under Contract DE-AC52-07NA27344. The work by GPB, DK and VIT was partly supported by the IARPA.